\documentclass[a4paper, 11pt, oneside]{article} 

\usepackage{geometry}
\usepackage{authblk}
\usepackage[utf8]{inputenc} 
\usepackage[T1]{fontenc} 
\usepackage{fouriernc} 
\usepackage{epsfig}
\usepackage{lineno,hyperref}
\usepackage{graphicx}
\usepackage{algorithm, algorithmic}
\usepackage{amssymb}
\usepackage{amsmath}
\usepackage{multicol}
\usepackage{float}
\usepackage{subfigure}
\usepackage{threeparttable}
\usepackage{cite}

\begin{document} 

\begin{titlepage} 

	\centering 
	
	\scshape 
	
	\vspace*{\baselineskip} 
	
	
	\rule{\textwidth}{1.6pt}\vspace*{-\baselineskip}\vspace*{2pt} 
	\rule{\textwidth}{0.4pt} 
	
	\vspace{0.75\baselineskip} 
	
	{\LARGE A veracity preserving model \\ for \\ synthesizing scalable electricity load profiles\\} 
	
	\vspace{0.75\baselineskip} 
	
	\rule{\textwidth}{0.4pt}\vspace*{-\baselineskip}\vspace{3.2pt} 
	\rule{\textwidth}{1.6pt} 
	
	\vspace{2\baselineskip} 
	
	
	
	\vspace*{3\baselineskip} 
	
	
	Edited By
	
	\vspace{0.5\baselineskip} 
	
	{\scshape\Large Yunyou Huang\\Jianfeng Zhan \\ Chunjie Luo \\ Lei Wang \\Nana Wang\\Daoyi Zheng\\Fanda Fan\\Rui Ren} 
	
	\vspace{0.5\baselineskip} 

	\vfill 
	
	
	\epsfig{file=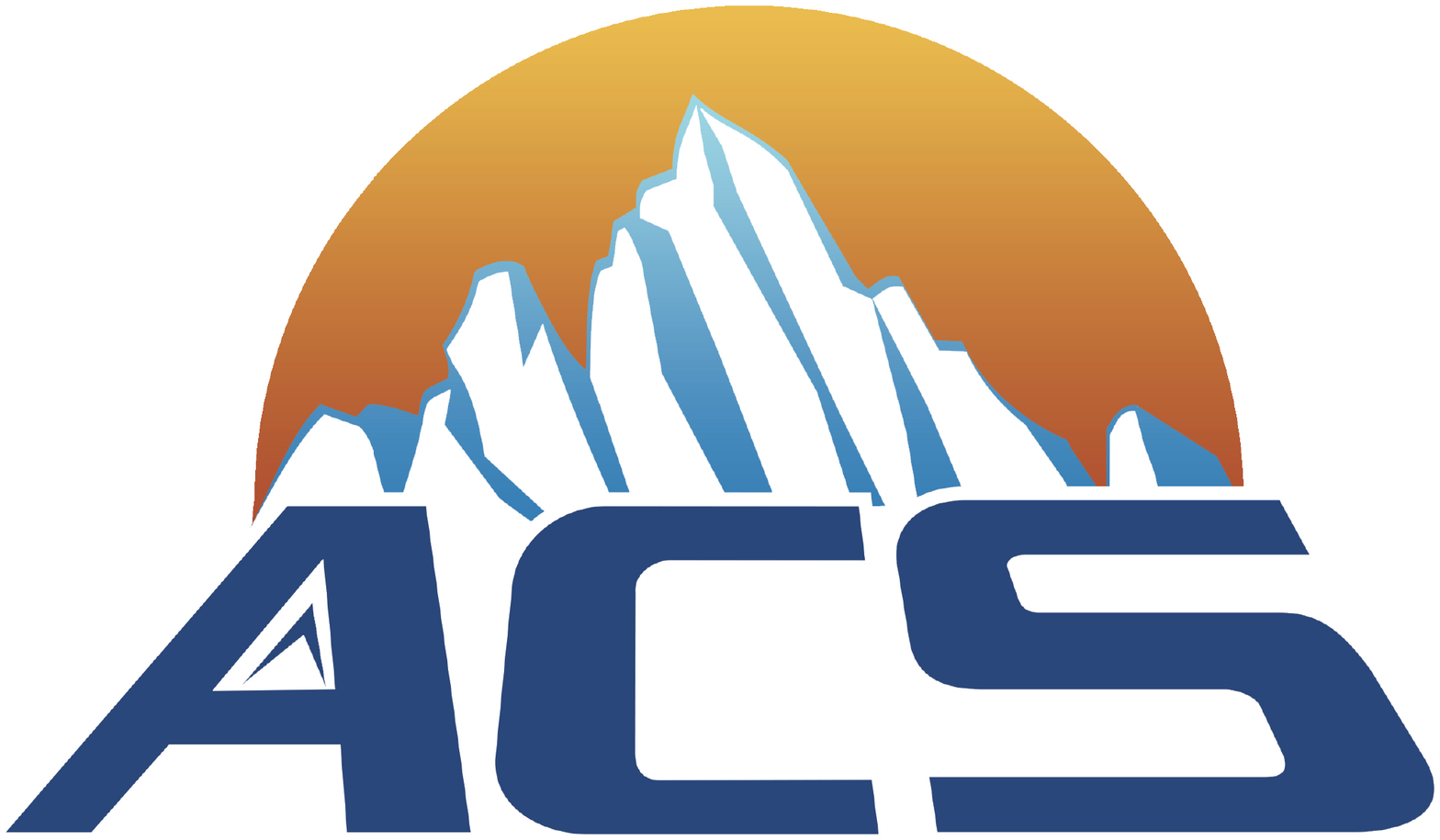,height=5cm,angle=270}
	\textit{\\Software Systems Laboratory (SSL), ACS\\ICT, Chinese Academy of Sciences\\Beijing, China\\http://prof.ict.ac.cn/ssl} 
	\vspace{5\baselineskip} 

	Technical Report No. ACS/SSL-2018-4 
	
	{\large February 8, 2018} 

\end{titlepage}


\title{A veracity preserving model for synthesizing scalable electricity load profiles}

\author[1,2,3]{Yunyou Huang}
\author[,1,2,3]{Jianfeng Zhan\thanks{zhanjianfeng@ict.ac.cn}}
\author[1,2,3]{Chunjie Luo}
\author[1,3]{Lei Wang}
\author[2]{Nana Wang}
\author[1,2,3]{Daoyi Zheng}
\author[1,2,3]{Fanda Fan}
\author[1,2,3]{Rui Ren}

\affil[1]{State Key Laboratory of Computer Architecture, Institute of Computing Technology(ICT), Chinese Academy of Sciences(CAS), Beijing 100080, China}
\affil[2]{University of Chinese Academy of Sciences, No. 19A Yuquan Road, Beijing 100049, China}
\affil[3]{Software Systems Laboratory, ACS, ICT, CAS, Beijing 100080, China}

\date{February 8, 2018}
\maketitle

\begin{abstract}
Electricity users are the major players of the electric systems, and electricity consumption is growing at an extraordinary rate. The research on electricity consumption behaviors is becoming increasingly important to design and deployment of the electric systems. Unfortunately,  electricity load profiles  are difficult to acquire. Data synthesis is one of the best approaches to solving  the lack of data, and the key is the model that preserves the real electricity consumption behaviors. In this paper, we propose a hierarchical multi-matrices Markov Chain (HMMC) model to synthesize scalable electricity load profiles that preserve the real consumption behavior on three time scales: per day, per week, and per year. To promote the research on the electricity consumption behavior, we use the HMMC approach to model two distinctive raw electricity load profiles. One is collected from the resident sector, and the other is collected from the non-resident sectors, including different industries such as education, finance, and manufacturing. The experiments show our model performs much better than the classical Markov Chain model. We publish two trained models online, and researchers can directly use these trained models to synthesize scalable electricity load profiles for further researches.
\end{abstract}


\clearpage

\section{Introduction}
Electric consumption is experiencing a rapid growth. The National Bureau of Statistics of the People's Republic of China reported that the electric consumption was 5503.21 billion KWH in 2015~\cite{NationalData}. Increasing energy efficiency is even more pressing. Understanding electricity consumption behaviors is the key to energy efficiency improvement. For example, in the Demand Response programs, the price strategy must base on the electricity consumption behaviors of users. Thus, the researches on electricity consumption behaviors have been attracting a lot of attention~\cite{8Viegas2016Classification,9Kwac2014Targeting,10Kwac2013Utility,11Bai2016Real,12Costa2016Inferring,13Hart1992Nonintrusive,14Mcloughlin2015A,15Ben2016Dynamic,16Kwac2013Utility,17Buitrago2016ELECTRIC,18Viegas2016Mining,19Shenoy2015Stochastic,20Li2015Research,21Yang2015Probabilistic}.

All of the researches on electricity consumption behaviors depend on the real electricity load profiles. However, obtaining the real electricity load profiles is non-trivial for most of researches. User privacy and commercial value of data are the main obstacles to acquiring electricity load profiles. There are few publicly available electricity load profiles. Moreover, they are small and only collected from the resident sector~\cite{22PecanStreetInc,23Pereira2014SustData,24Beckel2014The}.

Data synthesis is one of the best approaches to tackling  the lack of data, and the key is the model that preserves the real electricity consumption behaviors. The electricity consumption behavior is represented by the electricity load profile. In other words, the key of the data synthesis is preserving the real features of the electricity load profile. There are two difficulties in modeling electricity load profiles.
First, the electricity consumption model should preserve the features of real electricity load profiles, especially, the features on different time scales since the trend of change in the electricity load profiles varies in different periods.
Second, due to the huge differences among user activities, users have various electricity consumption habits. Thus, the electricity consumption model should contain as many types of user as possible, e.g., residential, industrial, and commercial users.
To the best of our knowledge, there is not an electricity consumption model that meets the requirements above.

\begin{figure*}
\centering
\includegraphics[width=6in]{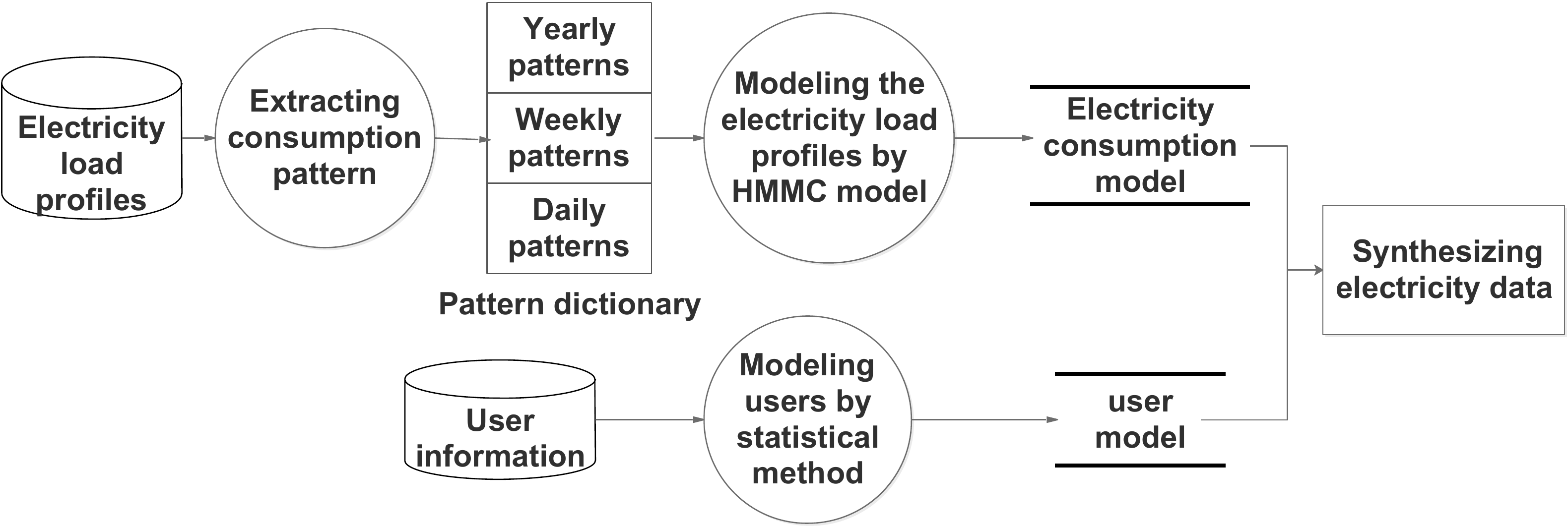}
\caption{The summary of our approach.\label{Fig1}}
\end{figure*}

In this paper, we propose a new approach to modeling electricity load profiles on the basis of the electricity consumption patterns shown in Figure~\ref{Fig1}. Our approach is as follows:  first, user electricity consumption patterns, each of which represents a class of similar electricity load profiles, are extracted from the electricity load profiles on three time scales: per day, per week, and per year. Second, a new hierarchical multi-matrices Markov Chain (HMMC, in short) model is proposed to characterize the electricity load profiles in terms of electricity consumption patterns. The model is separated into three levels: per year, per week, and per day accordingly. For every pattern of different time scales, a multi-matrices Markov Chain (MMC) model is used to represent it. Meanwhile, the user is modeled by a statistical method. Third, the user model and the HMMC model are used to synthesize the scalable electricity load profiles for researches. Finally, the proposed method is evaluated on a smart grid community data set provided by Pecan Street Inc~\cite{22PecanStreetInc}. The experiments shows that  our method can effectively preserve the real features of electricity load profiles on different time scales in comparison with the classic Markov Chain model.

To promote the research on the electricity consumption behaviors, we use the HMMC model to characterize two real electricity load profiles. One  data set is provided by Pecan Street Inc., containing more than 800 users' data, which is collected from the resident sector. The other one is a confidential data set\footnote{The data is collected from the non-resident sectors. we have obtained
permission to use. The name cannot be disclosed per the non-disclosure agreements
we have signed.}, containing more than 80000 users' data, which is collected from different industries such as education, finance and manufacturing. We publish the two trained models online \footnote{The models and generator will be published on~\url{http://prof.ict.ac.cn/} soon.}. Researchers can directly use these trained models to synthesize scalable electricity load profiles.

The paper is structured as follows. Section 2 summarizes the publicly available dataset and the previous work on modeling data. Section 3 proposes our model. Section 4 presents and discusses the results. Section 5 describes the trained models of the electricity consumption data. Section 6 draws a concluding remark.

\section{Related Work}
It is very detrimental to performing research on the electricity consumption behaviors without publicly available data sets. Like other domains, such as natural language processing and image processing, the public availability of data sets is  fundamental in improving related techniques. There are few public available data sets in the energy field. Pecan Street Inc Dataport2017 is a publicly available electricity consumption data set. It involves more than 800 users since 2011~\cite{22PecanStreetInc}. Collected from 50 families~\cite{23Pereira2014SustData}, SustData is a free data set, containing power usage and related user information. The ECO data set is collected from 6 households in Switzerland over a period of 8 months (From June 2012 to January2013)~\cite{24Beckel2014The}.

Most of the publicly available electric data sets are collected for Non-intrusive load monitoring (NILM) research. They provide different internal load compositions and the user's related information. On the overall, these data sets contain very few users. Moreover, the public electricity data sets are collected only from the resident sector, and hence insufficient for the research on electricity consumption behaviors. For example, power grid planning need to take as many users of different types as possible into consideration.

Data synthesis is an approach to solving the lack of data. The generative model attractes a lot of attention and a number of models have been proposed in different areas. In the field of marketing, the mixture of normal distributions provides a useful extension of the normal distribution for modeling daily changes in market variables~\cite{wang2001generating}. In the field of solar radiation, the Markov model is used to model and generate the global solar radiation data~\cite{ngoko2014synthetic}. In the field of wind power, the chronological or sequential Monte Carlo simulation is applied to model and synthesize wind power time series~\cite{mosayebian2016synthetic}.
However, the model on the electricity consumption behaviours is rare in the field of energy. Duffy~\cite{30Duffy2010The} develops a generative model---the first order Markov Chain model---to model the electricity load profiles for five individual dwelling types in Ireland. However, the first-order Markov Chain model can not effectively preserve the real features of the user electricity load profiles as shown in Section 4.

This paper fills the research gap mentioned above. We propose an innovative model to synthesize scalable electricity load profiles that contains different types of electric users. Moreover, the scalable data set, synthesized using our approach, preserves the real user electricity consumption behaviors.

\section{Methodology}
Our methodology is as follows: First, we separate the original data into electricity load profiles, and the user data, which represents the user information. Second, given the electricity load profiles, we extract the electricity consumption patterns on different time scales using the clustering algorithm. Third, we propose the HMMC model to model the electricity load profiles. The modeling of the electricity load profiles is separated into three levels: per year, per week, and per day. For each pattern of different time scales, an MMC model is used to represent it. For the user data, it is modeled using a statistical method.

\subsection{The electricity consumption model}
\subsubsection{Extracting the consumption patterns}

\begin{figure*}
\centering
\includegraphics[width=6in]{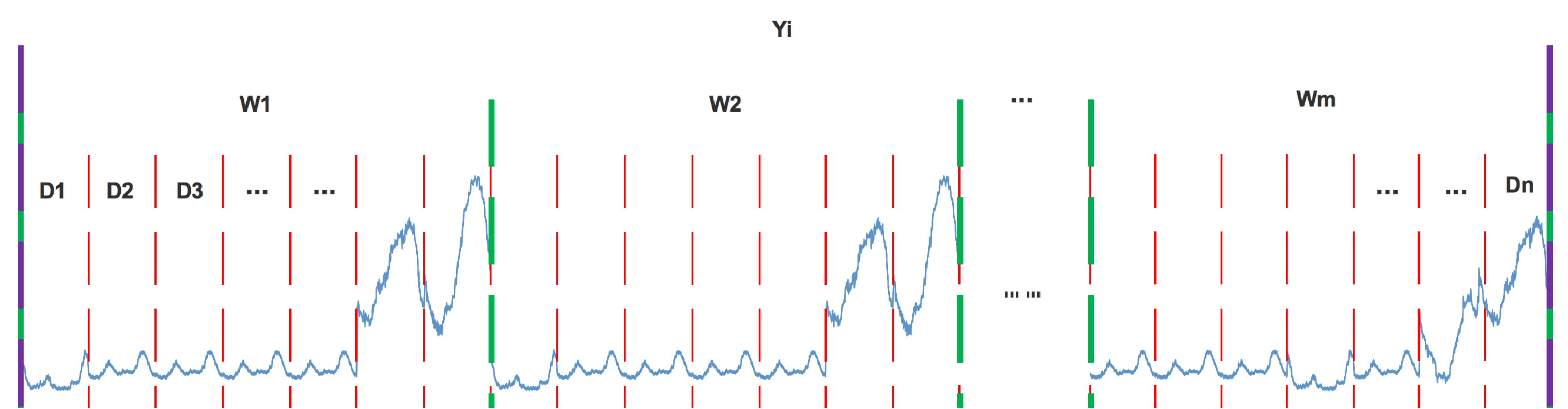}
\caption{Obtaining per-year segments, per-week segments and per-day segments from the user electricity load profile. The $Y_i$ are the yearly electricity load profiles, the $W_m$ are the weekly electricity load profiles and the $D_n$ are the daily electricity load profiles. ~\label{Fig2}}
\end{figure*}

In order to extract the electricity consumption patterns of different time scales, the electricity load profiles are divided into segments. As shown in Figure~\ref{Fig2}, each electricity load profile is divided on three
time scales: per year, per week, and per day. Segments $Y_i$ are obtained after the user electricity load profile is divided per year. Likewise, Segments $W_m$ are obtained after the user electricity load profile is divided per week, and segments $D_n$ are obtained after the user electricity load profile is divided per day, respectively. After the electricity load profiles of all the users are handled on the above, we obtain the yearly, weekly, and daily  electricity load profile sets of all users:  $S_y=\{Y_i\}$,  $S_w=\{W_m\}$, $S_d=\{D_n\}$. And the yearly, weekly, and daily patterns are  extracted from the corresponding electricity load profile sets using the following clustering algorithm, respectively.

Due to the simplicity and low time complexity of the K-Means, a modified K-Means algorithm, called adaptive K-Means,  is used to extract the electricity consumption patterns. In order to ensure the similarity of the profiles within the same cluster, two metrics: the standard deviation of the total consumption $\sigma_{total}$ and the mean of the total consumption $\mu_{total}$ are used  in the adaptive K-Means. As shown in  Algorithm~\ref{alg1}, for each cluster $C_i$ of the K-Means' result, if $\frac{\sigma_{total\_i}}{\mu_{total\_i}}$ is bigger than a threshold $\gamma$, the K-Means algorithms is iterated on  $C_i$ until $\frac{\sigma_{total\_x}}{\mu_{total\_x}}$ in every cluster is smaller than $\gamma$. Our approach can automatically determine the number of $K$.
\begin{algorithm}[!htbp]
         \renewcommand{\algorithmicrequire}{\textbf{Input:}}
         \renewcommand{\algorithmicensure}{\textbf{Output:}}
         \caption{The Adaptive K-Means algorithm}
         \label{alg1}
         \begin{algorithmic}[1]
                   \REQUIRE The electricity load profiles of all users $DS=\{s_n (t)\}$, Initial and max number of clusters ($K_{initial}$, $K_{max}$) and $\gamma$.
                   \ENSURE The electricity load profiles with Cluster id $C=\{SC_i=\{<id,s_n (t)>\}\}$, cluster centers $Cnt=\{c_k(t)\}$.
                   \STATE Set $K= initial\_k$ , $Cnt=Cnt_{tmp}=\varnothing$ , $C=\varnothing$ and $stop=false$
                   \WHILE {$(K<max\_k ) and (!stop)$}
                   \STATE $stop=true$
                   \STATE Run K-Means with $K$ and $Cnt_{tmp}$ for $DS$. Output the clusters $C$ and cluster centers $Cnt$.
                   \STATE $Cnt_{tmp}$=$Cnt$
                   \FORALL{$ SC_i \in C $}
                   \IF {$\frac{\sigma_{total\_i}}{\mu_{total\_i}}>\gamma$}
                   \STATE $stop=false$
                   \STATE Run K-Means with $2$ and $\varnothing$ for $SC_i$. Output the clusters $C_{SC_i}$ and cluster centers $Cnt_{SC_i}$.
                   \STATE $Cnt_{tmp}=Cnt_{tmp}\cup Cnt_{SC_i}$
                   \ENDIF
                   \ENDFOR
                   \STATE $K$=size($Cnt_{tmp}$)
                   \ENDWHILE
                   \STATE \textbf{return} $Cnt$, $C$
         \end{algorithmic}
\end{algorithm}

\subsubsection{Model the consumption patterns }
Markov Chain is a stochastic process with Markov properties. In Markov Chain, each state only depends on the immediately preceding $l$ states. In this paper, when $l=1$ we call it classic or first-order Markov Chain. When $l>1$ we call it $lth$-order Markov Chain. And the whole of the immediately preceding $l$ states is called a preceding subsequence.

\begin{figure}
\centering
\includegraphics[width=3in]{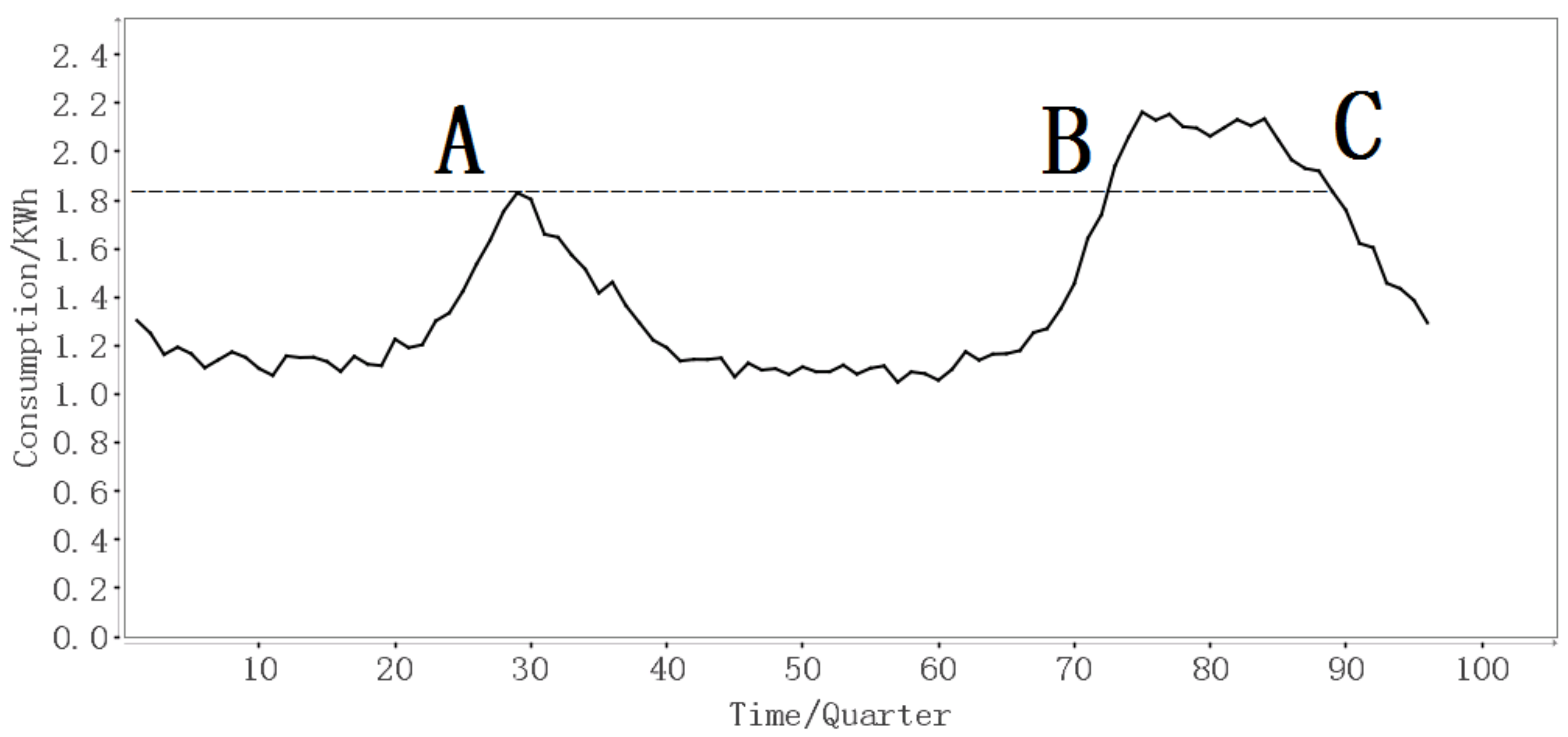}
\caption{Dual peak morning \& afternoon electricity load profiles.~\label{Fig3}}

\end{figure}

For the classical Markov Chain model, there are two defects in modeling electricity consumption pattern.
First, each subsequent state depends only on the immediately preceding one. It does not care about the position of the state in the sequence. Thus, the same preceding state changes into the same current state in different positions of the sequence only with one probability. However, the position in the sequence---timing  in the daily life---has huge impact on electricity consumption behavior.
 For example, in the morning and afternoon, though there are some states having the same preceding subsequences, due to user habits the preceding subsequences usually change into the same subsequent state with different probabilities.
For instance, Figure~\ref{Fig3} shows an electricity load profile,  which belongs to the classic Dual Peak Morning \& Afternoon electricity consumption pattern. Using the first-order Markov Chain, $A$, $B$ and $C$ are considered as the same preceding subsequences. However, they not only tend to change into different subsequent states, but also change into the same subsequent state with different probabilities. When $l<4$ $*A$\footnote{The $*$ represents $l-1$ preceding states of $A$.} and $*B$  are the same preceding subsequences, using the higher-order Markov Chain can model both the cases that the same preceding subsequences either change into different subsequent states or the same subsequent state with different probabilities.

\begin{figure}[!htbp]
\centering
\includegraphics[width=3in]{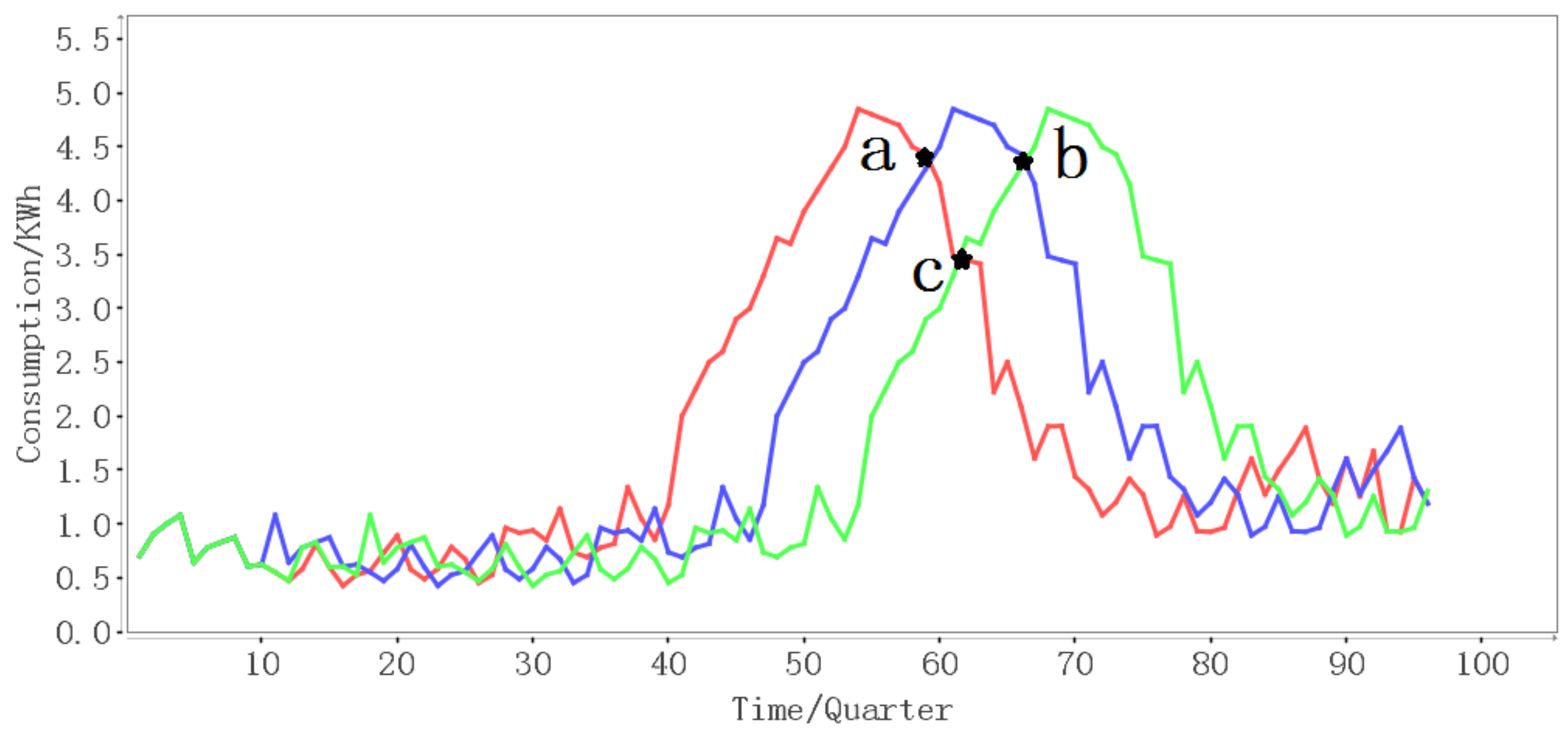}
\caption{Three electricity load profiles in the same pattern: A, B, and C. A is red, B is blue, and C is green.~\label{Fig4}}
\end{figure}

The other defect is that the differences between the electricity load profiles within the electricity consumption pattern can be accumulated when the new electricity load profile is being synthesized by the classic Markov Chain. In a particular pattern, electricity load profiles  have differences, which can be accumulated. Figure~\ref{Fig4} shows an electricity consumption pattern with 3 electricity load profiles. When the pattern is modeled by the first-order Markov Chain, we may synthesize a new electricity load profile consisting of 3 parts: the first part is the part of $A$ before point $a$; the second part is the part of $B$ between points $a$ and $b$; and the last part is the part of $C$ after point $b$). Obviously, the new electricity load profile has significant difference from the raw data as shown in~\ref{Fig4}. As a result, when we synthesize the consumption data, we may obtain unreasonable consumption behavior.


To solve these problems mentioned above, we propose two methods. The first one is to use the higher-order Markov Chain to model the electricity consumption pattern. When $l$ is large enough, the same preceding subsequence changes into the same subsequent state in different positions of the sequence only with a probability. Thus, we can ignore the position of the state in the sequence. However, the transition probability matrix size grows exponentially with $l$. The second one is to use the multi-matrices Markov Chain (MMC) to model the electricity consumption pattern. MMC consists of many individual transition probability matrices, which are created for each adjunct point of the sequence. Thus, for every position of the sequence, a particular preceding subsequence changes into a particular subsequent state with a specific probability. Meanwhile it was found that MMC with a small $l$ can control the accumulation of differences between the electricity load  profiles within the pattern when we synthesize the new electricity load profile. According to Equation~\ref{eq1}, multiple transition probability matrices for $lth$-order Markov Chain with $n$ states are expressed as:
\begin{equation}\label{eq1}\centering
  p_i^k=
\left[
  \begin{array}{cccc}
    p_{1,1,...,1,1} & p_{1,1,...,1,2} &... & p_{1,1,...,1,n}\\
    p_{1,1,...,2,1} & p_{1,1,...,2,2} &... & p_{1,1,...,2,n}\\
    ... & ... & ... & ...\\
    p_{1,1,...,n,1} & p_{1,1,...,n,2} &... & p_{1,1,...,n,n}\\
    ... & ... & ... & ...\\
    p_{n,n,...,n,1} & p_{n,n,...,n,2} &... & p_{n,n,...,n,n}\\
  \end{array}
\right]
\end{equation}
Here, $k$ is the $ID$ of the electricity consumption pattern and $i$ is the position in the time series.

\subsubsection{Model the electricity load profiles}


In the data set we investigate, the electricity consumption data is collected every 15 minutes. It makes the yearly electricity load profile a high dimensional sequence. To model the corresponding pattern of the high-dimension yearly electricity load profiles with the MMC, we face two challenges.

First, when we use MMC to model the electricity consumption pattern, the number of transition probability matrix increases with the dimension of electricity load profile. For a yearly electricity load profile, the number of dimension is usually very large. For instance, the dimension number of a yearly electricity load profile is 35040 when the smart meter collects the load profile every 15 minutes, and hence MMC with 35039 transition probability matrices will be used to model a yearly consumption pattern. When the data sampling frequency of smart meters becomes more higher, we will get much higher-dimension yearly electricity load profiles, and the number of the transition probability matrix in MMC will be much larger accordingly.

\begin{figure}[!htbp]
\centering
\includegraphics[width=3in]{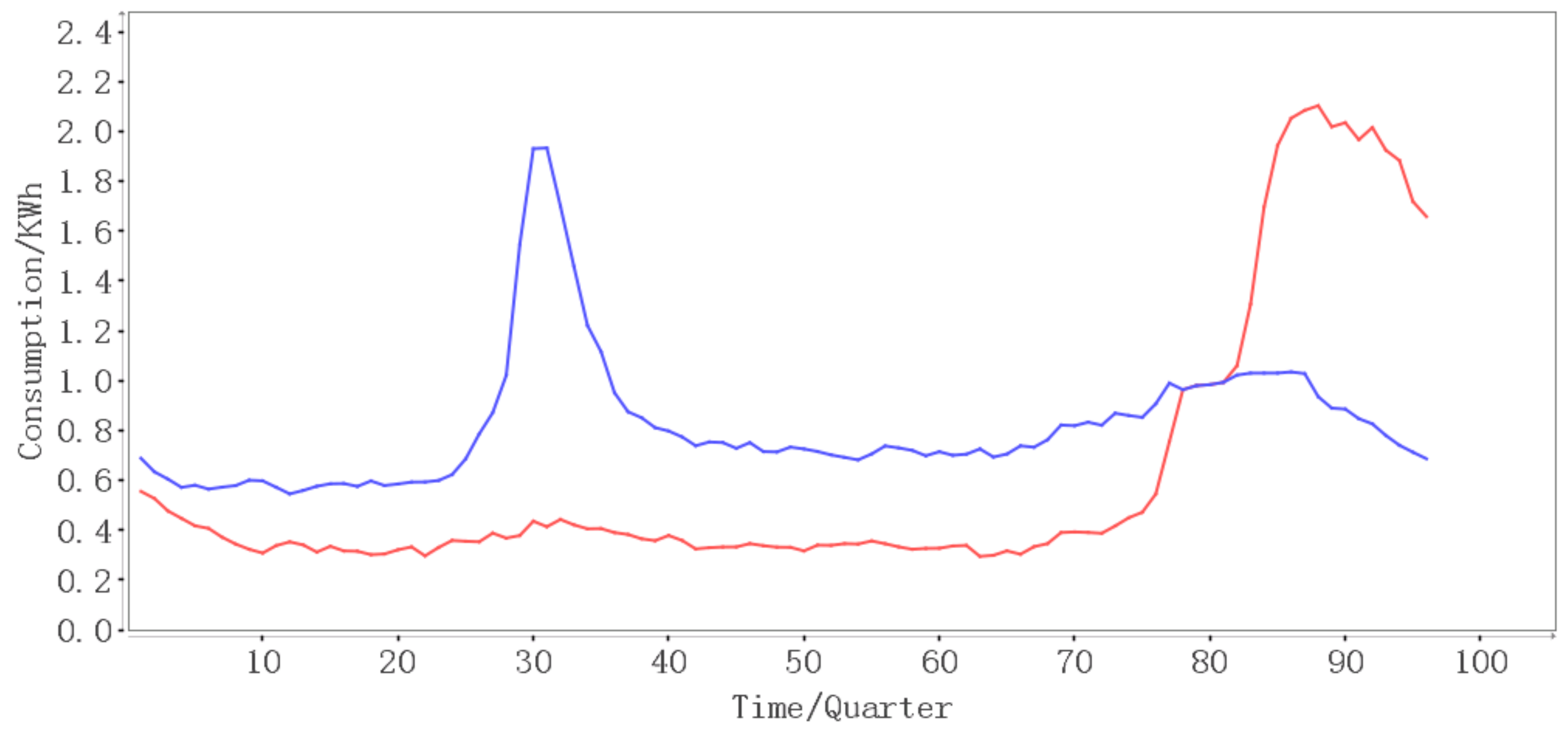}
\caption{A part of a yearly electricity load profile: there is only one peak for each pattern.~\label{Fig5}}

\end{figure}

\begin{figure}[!htbp]
\centering
\includegraphics[width=3in]{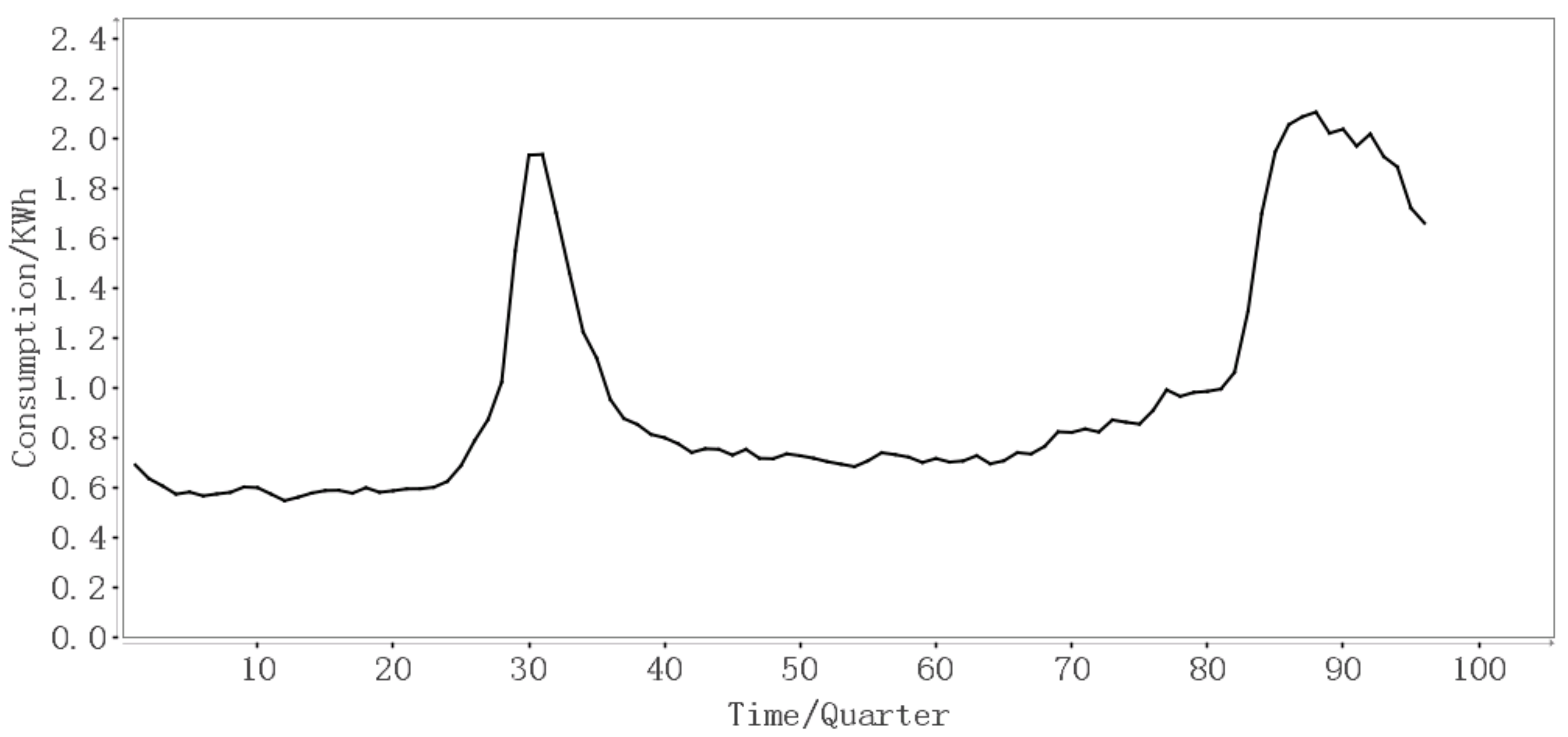}
\caption{Due to the fact that two  electricity profiles have intersection shown in Figure 5, we may synthesize electricity load profile with two peaks after using the MCC model to model the electricity load profiles.~\label{Fig6}}
\end{figure}

The second challenge is that the unreasonable part of the electricity load profiles may be synthesized when two significantly different  profiles co-exist  within a yearly pattern. For example, Figure~\ref{Fig5} shows that the red curve and blue one are two significantly different daily electricity load profiles at the same date.  When we synthesize the electricity load profile using the MMC model, it may generate a unreasonable part of  electricity load profile. Figure~\ref{Fig6} shows a negative example: there are two peaks at the synthesized electricity load profile while for the original data, there is only one peak for each load profile as shown in~\ref{Fig5}.


\begin{figure}[!htbp]
\centering
\includegraphics[width=3in]{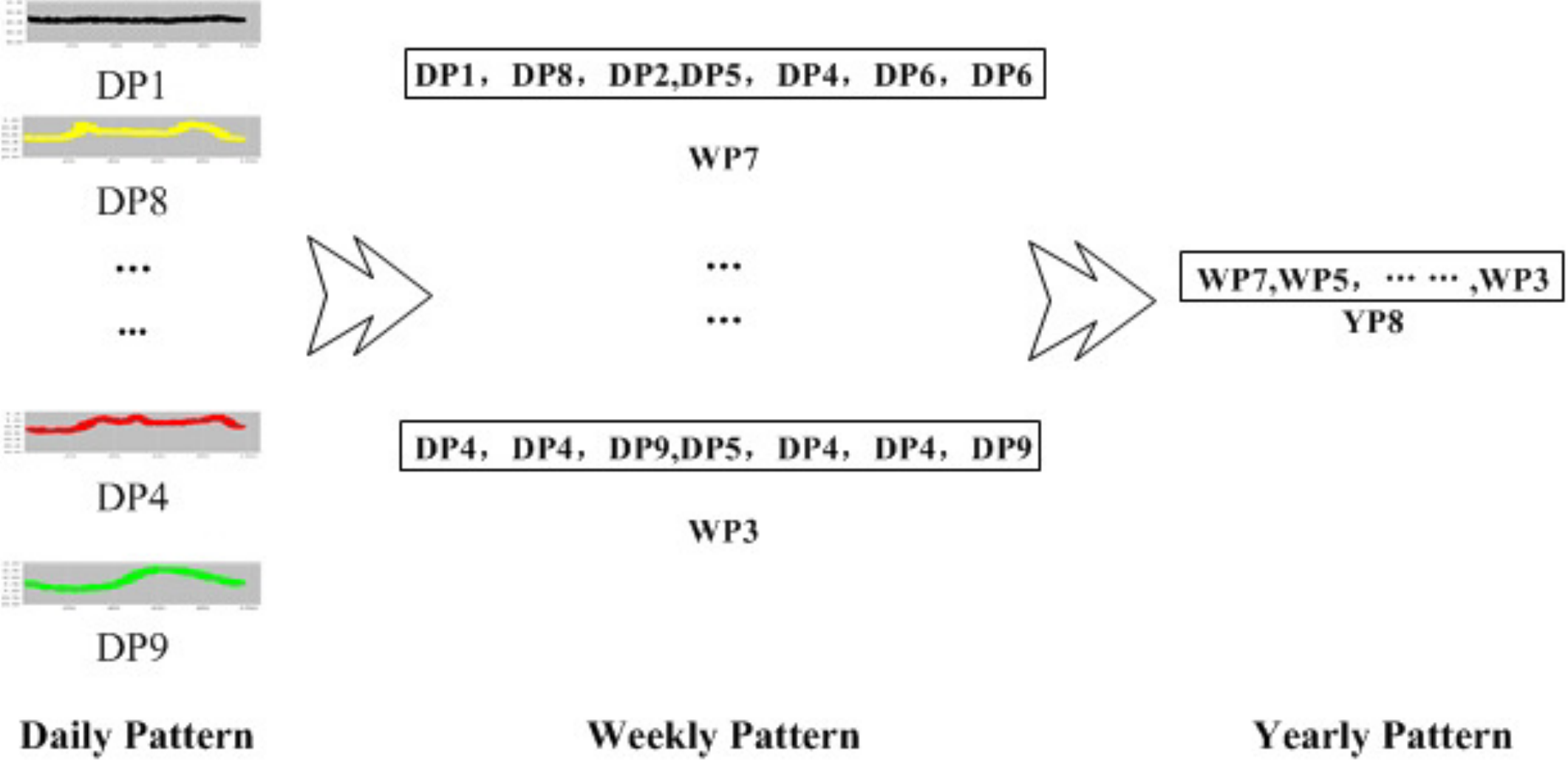}
\caption{Our bottom-up approach to synthesizing yearly electricity consumption patterns on the basis of weekly and daily electricity consumption patterns.~\label{Fig7}}

\end{figure}
\begin{figure}[!htbp]
\centering
\includegraphics[width=3in]{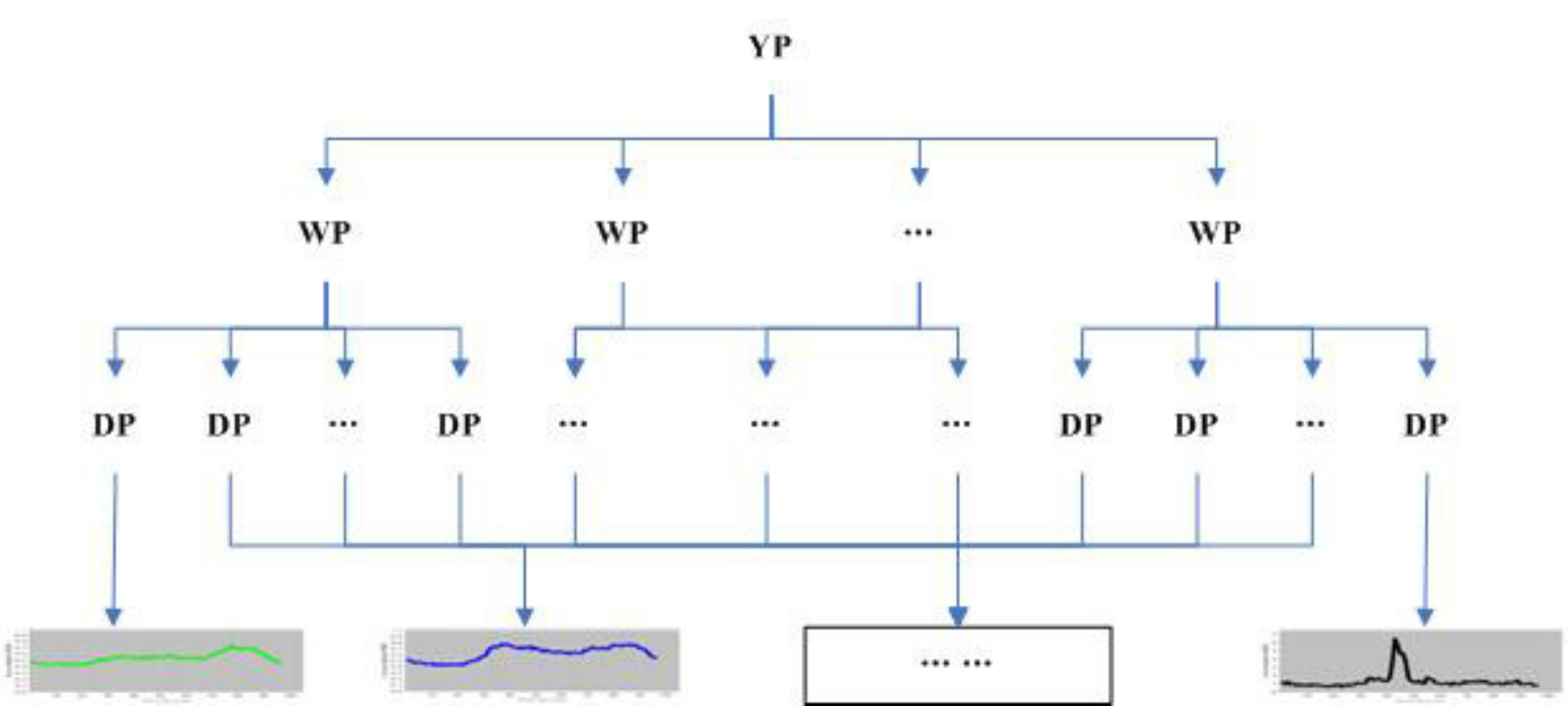}
\caption{The procedures of synthesizing electricity load profiles.~\label{Fig8}}
\end{figure}

\begin{figure*}[t]
\centering
\subfigure[The daily patterns]{\includegraphics[width=2in, height=0.8in]{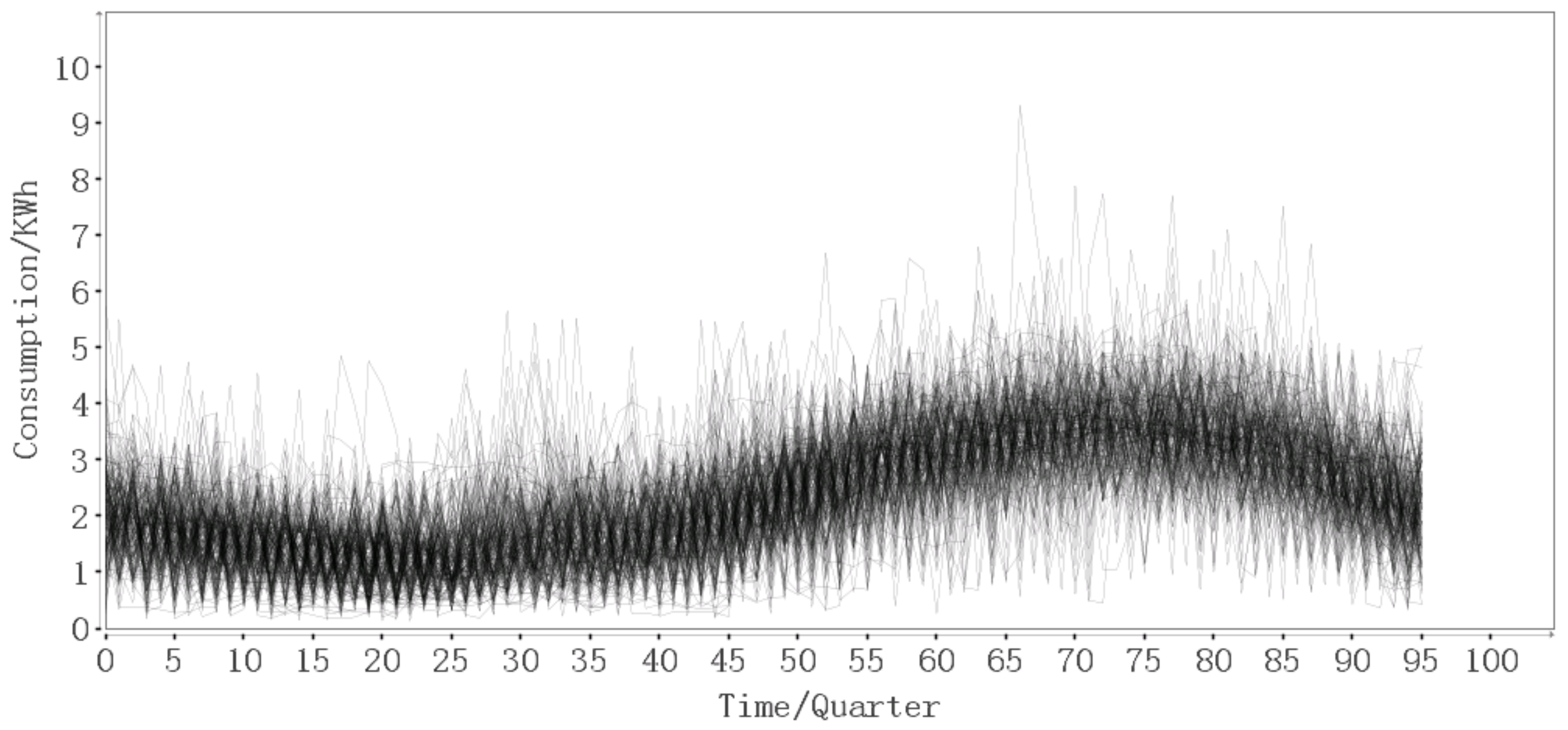}
\includegraphics[width=2in, height=0.8in]{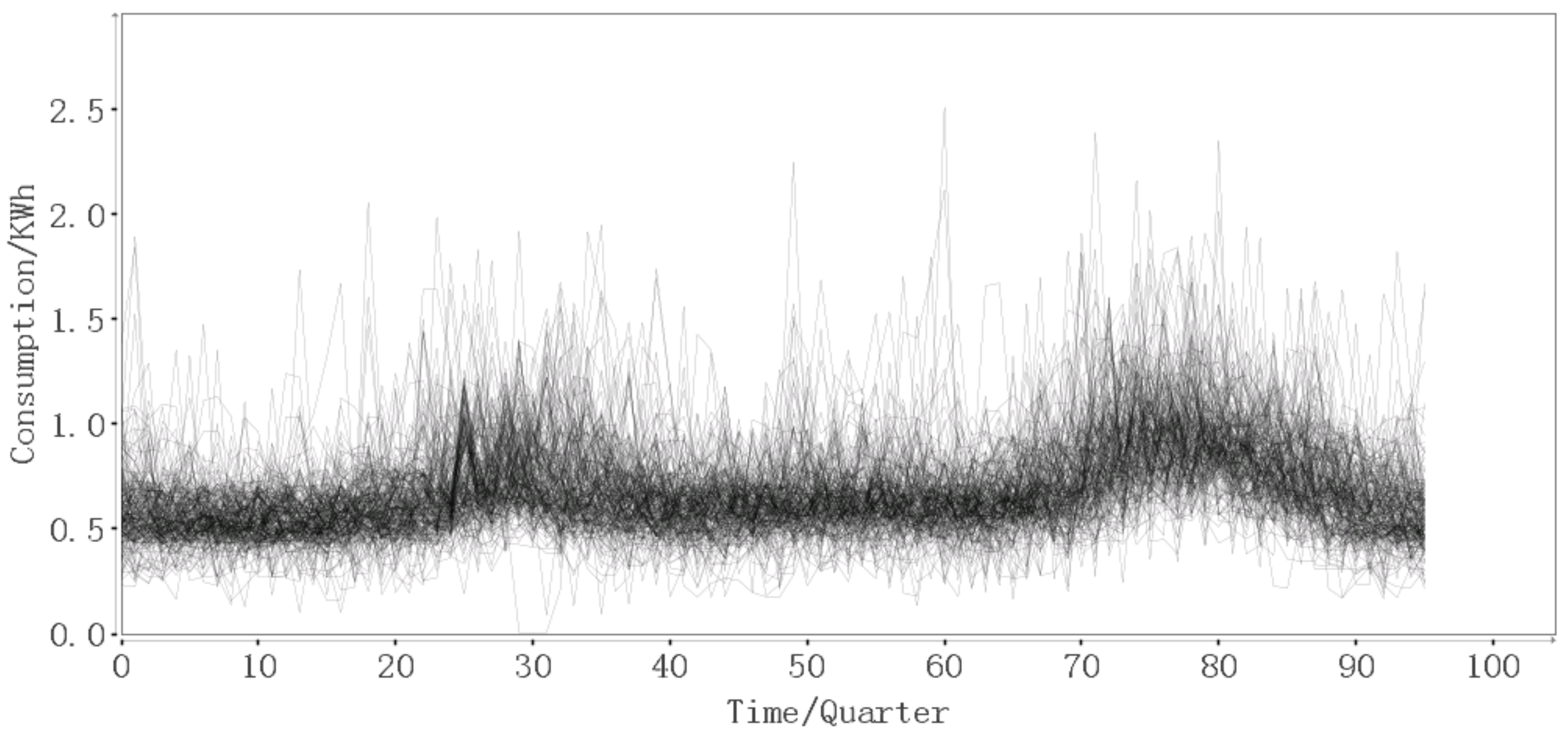}\includegraphics[width=2in, height=0.8in]{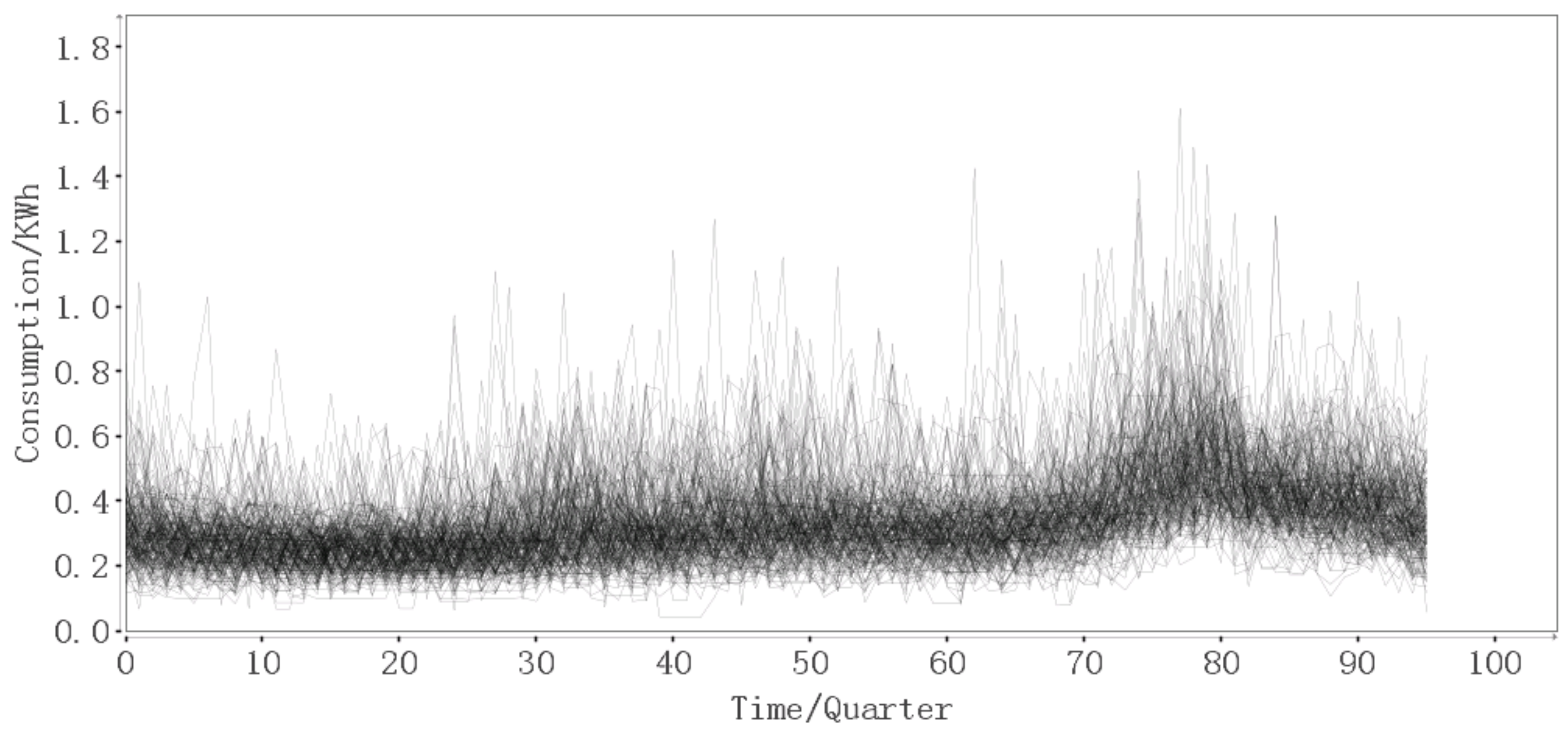}
}
\subfigure[The weekly patterns]{\includegraphics[width=2in, height=0.8in]{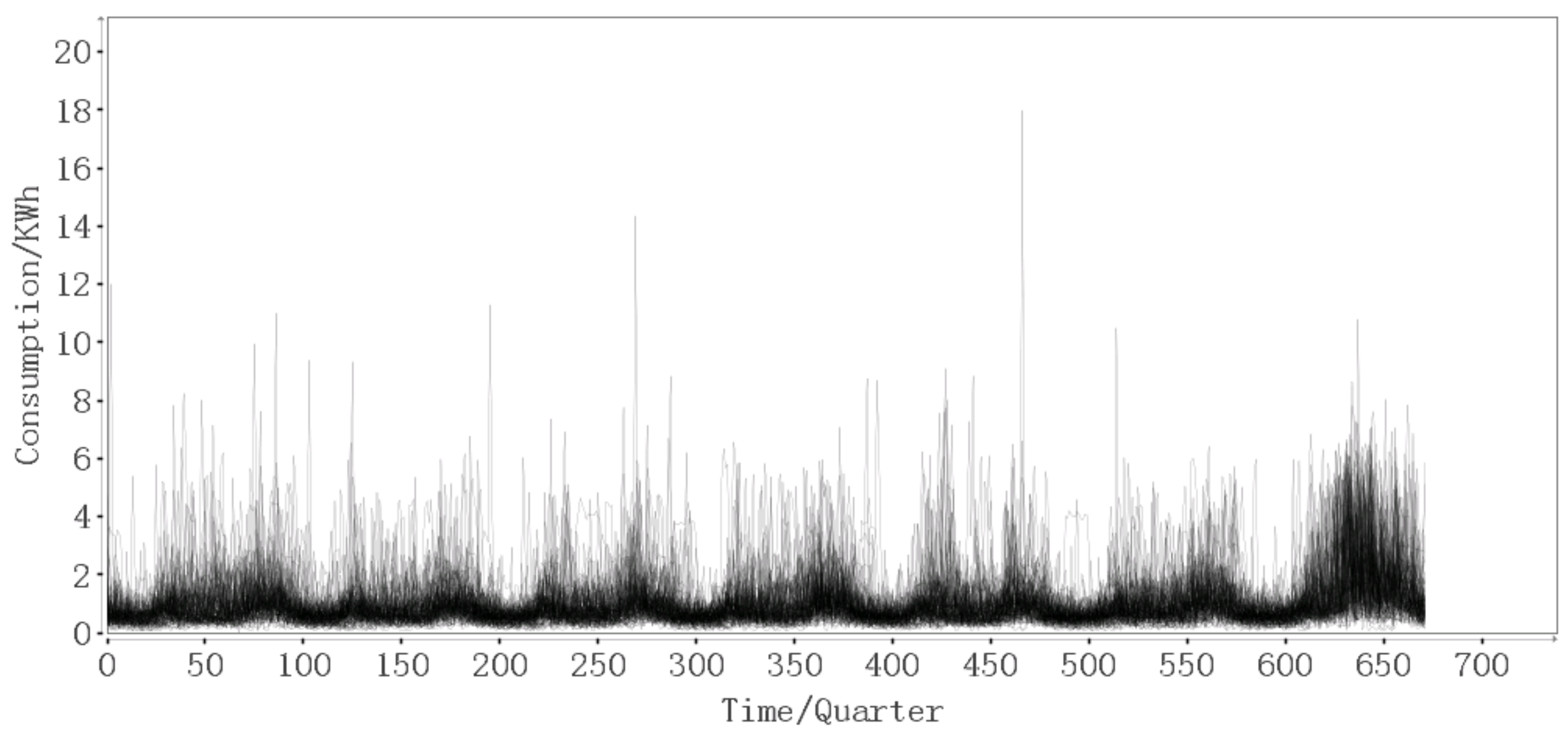}
\includegraphics[width=2in, height=0.8in]{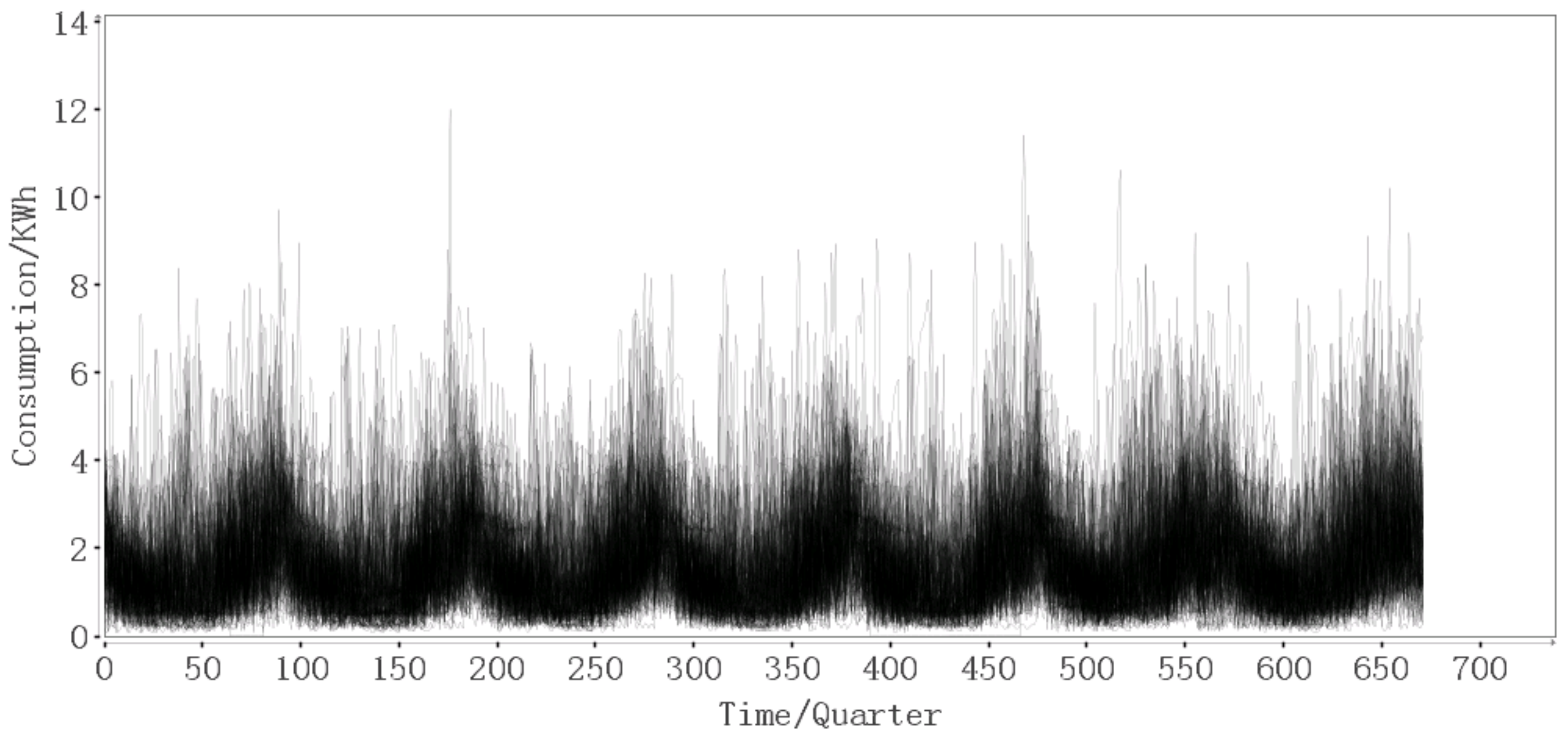}\includegraphics[width=2in, height=0.8in]{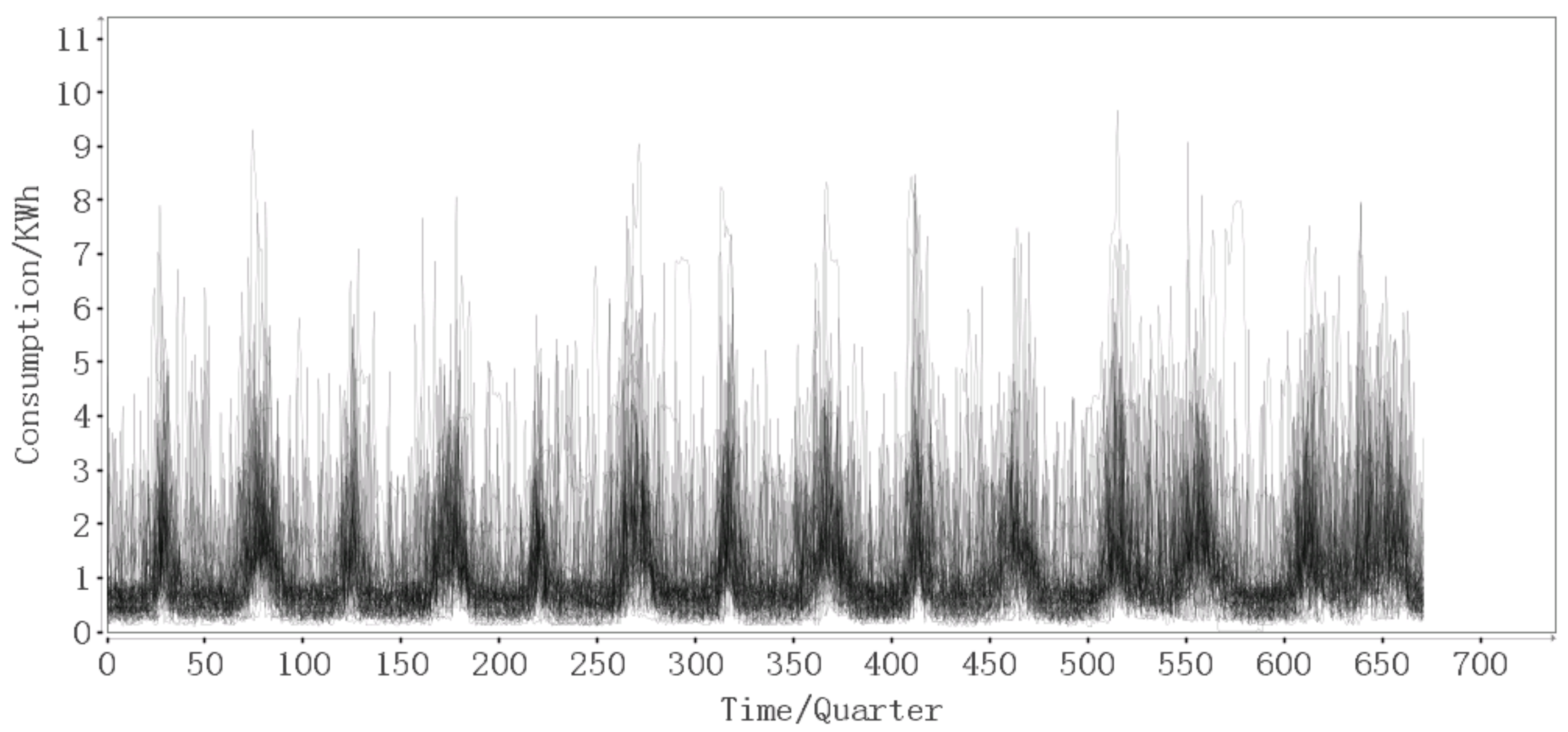}}
\subfigure[The yearly patterns]{\includegraphics[width=2in, height=0.8in]{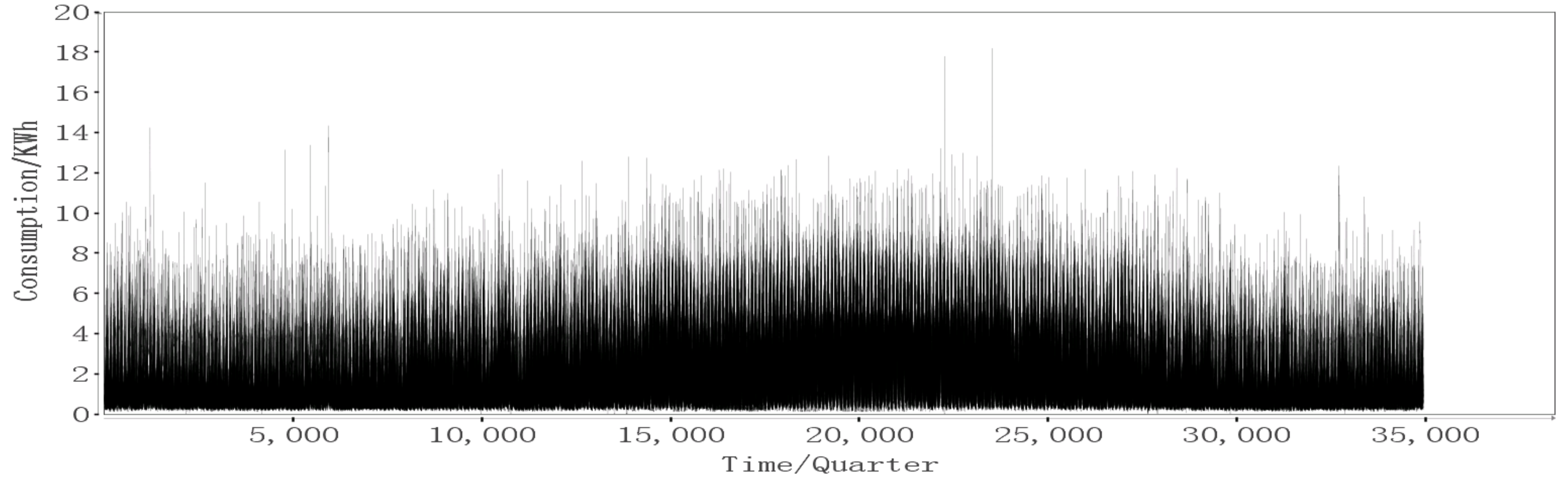}
\includegraphics[width=2in, height=0.8in]{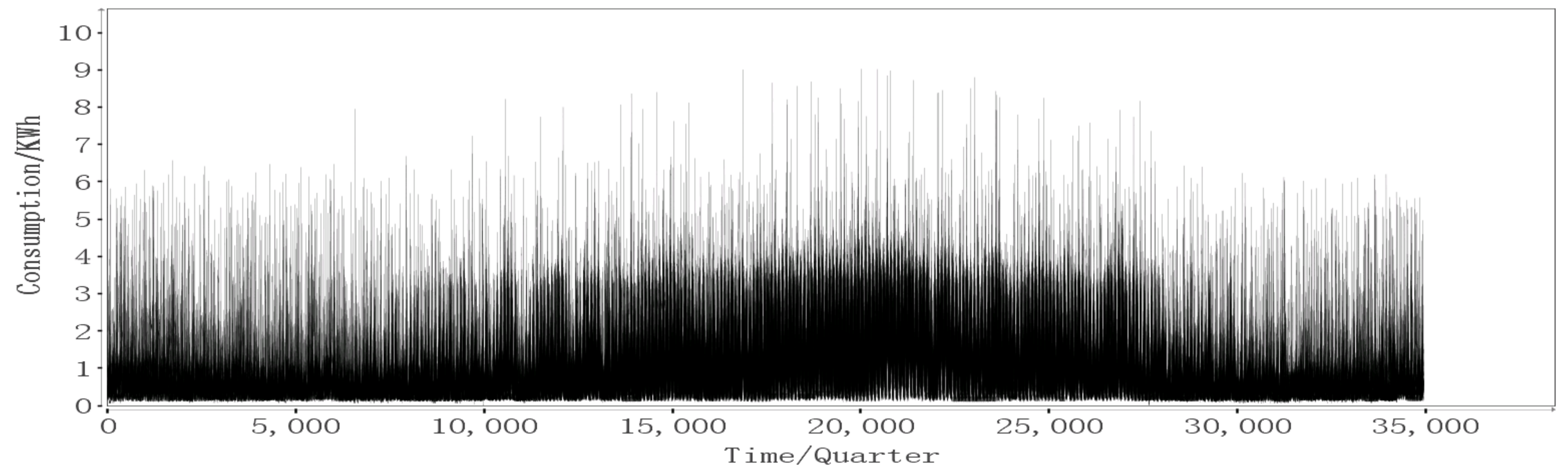}\includegraphics[width=2in, height=0.8in]{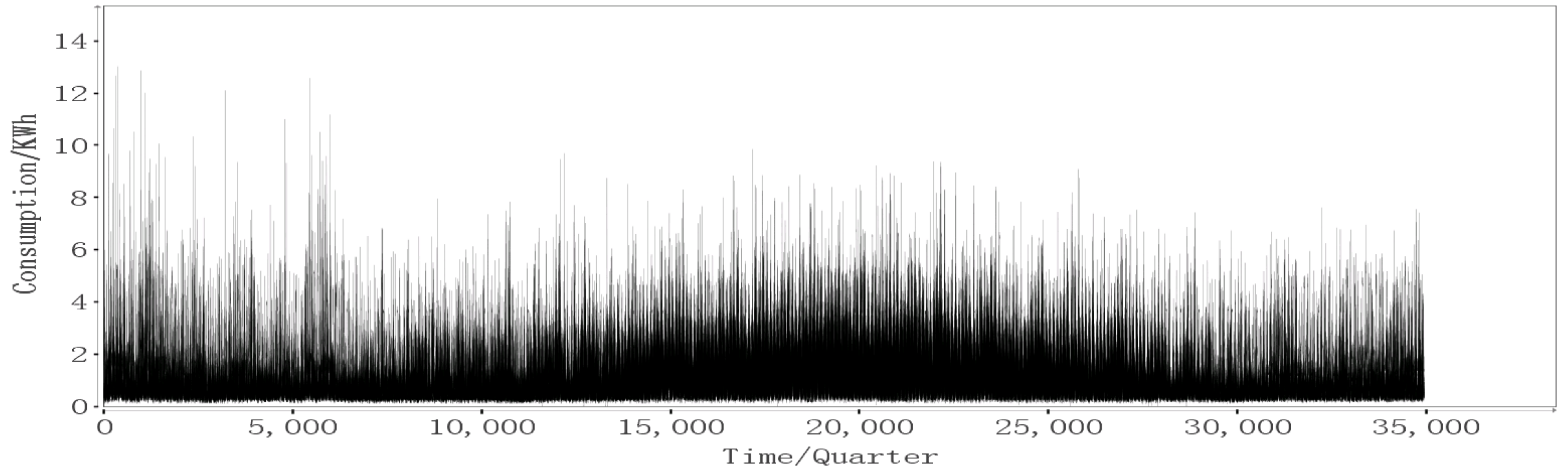}}
\caption{The pattern samples of  different time scales.\label{Fig9}}
\end{figure*}

\begin{figure*}[t]
\centering
\subfigure[The raw data.]{\includegraphics[width=2in, height=0.8in]{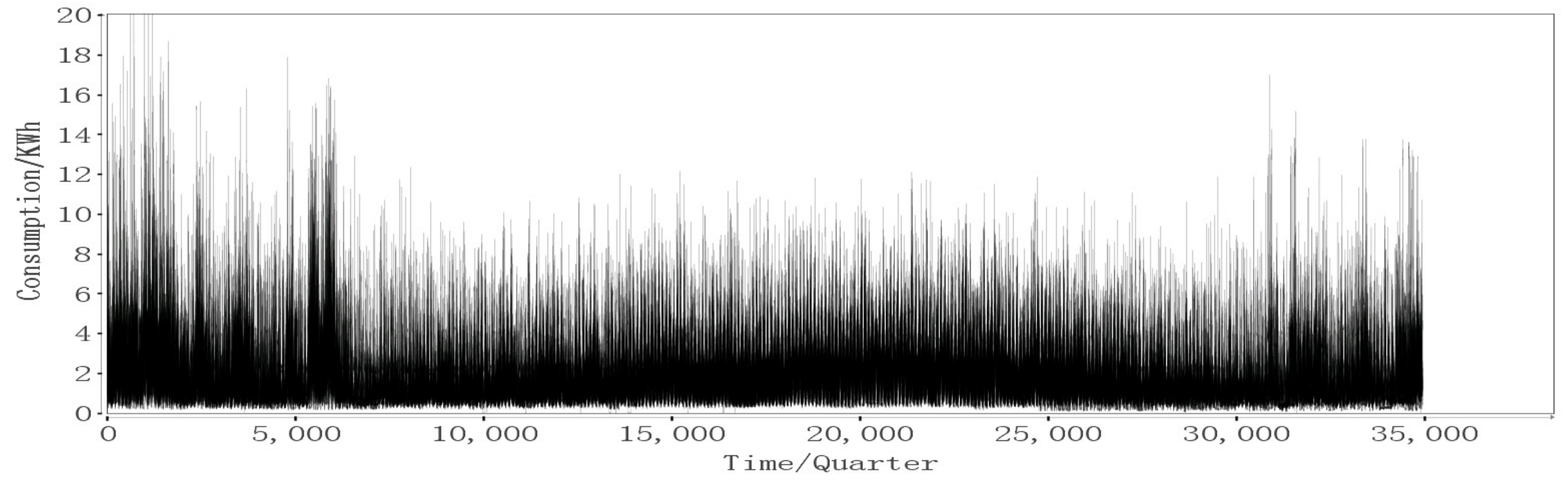}
\includegraphics[width=2in, height=0.8in]{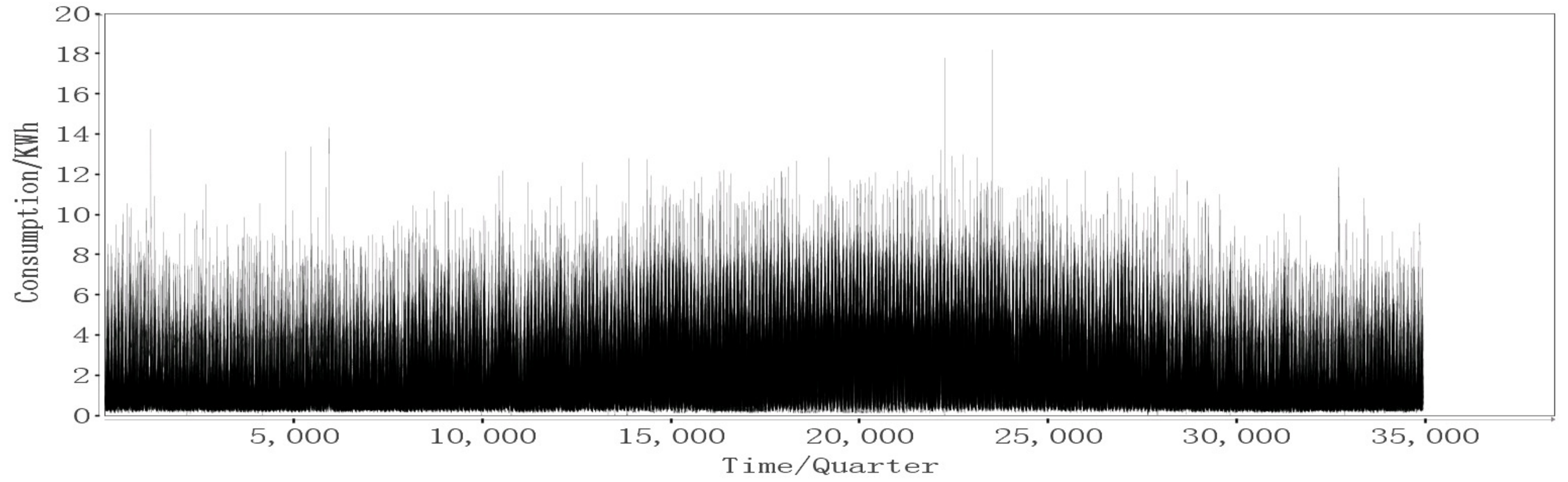}\includegraphics[width=2in, height=0.8in]{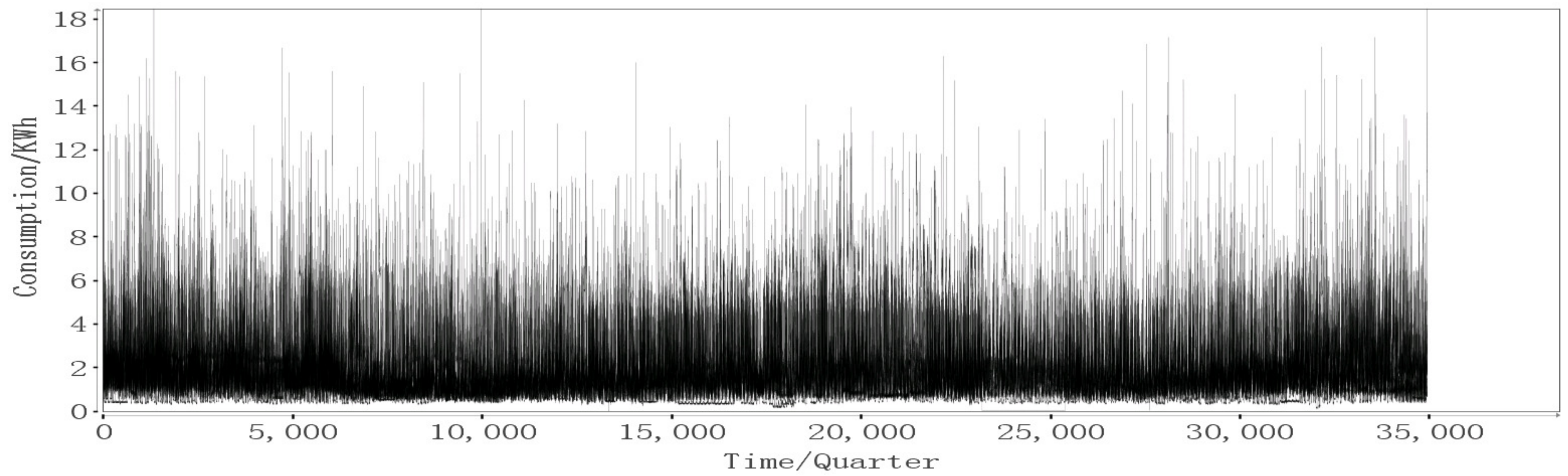}}
\subfigure[The data synthesized using the HMMC model trained with the raw data.]{\includegraphics[width=2in, height=0.8in]{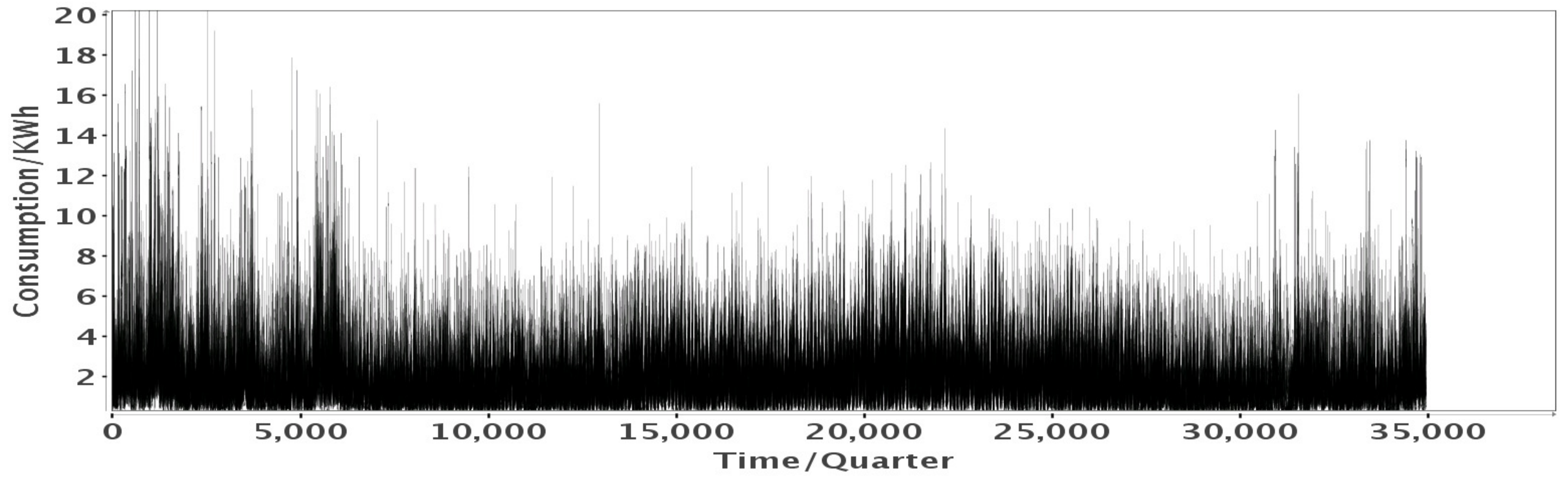}
\includegraphics[width=2in, height=0.8in]{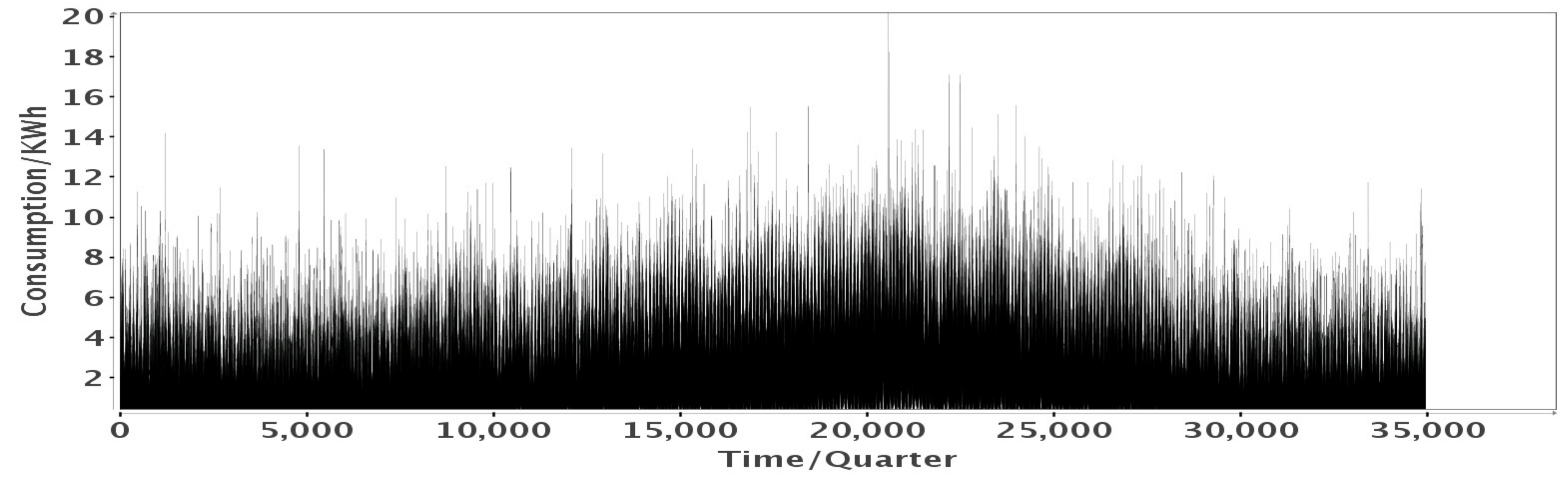}\includegraphics[width=2in, height=0.8in]{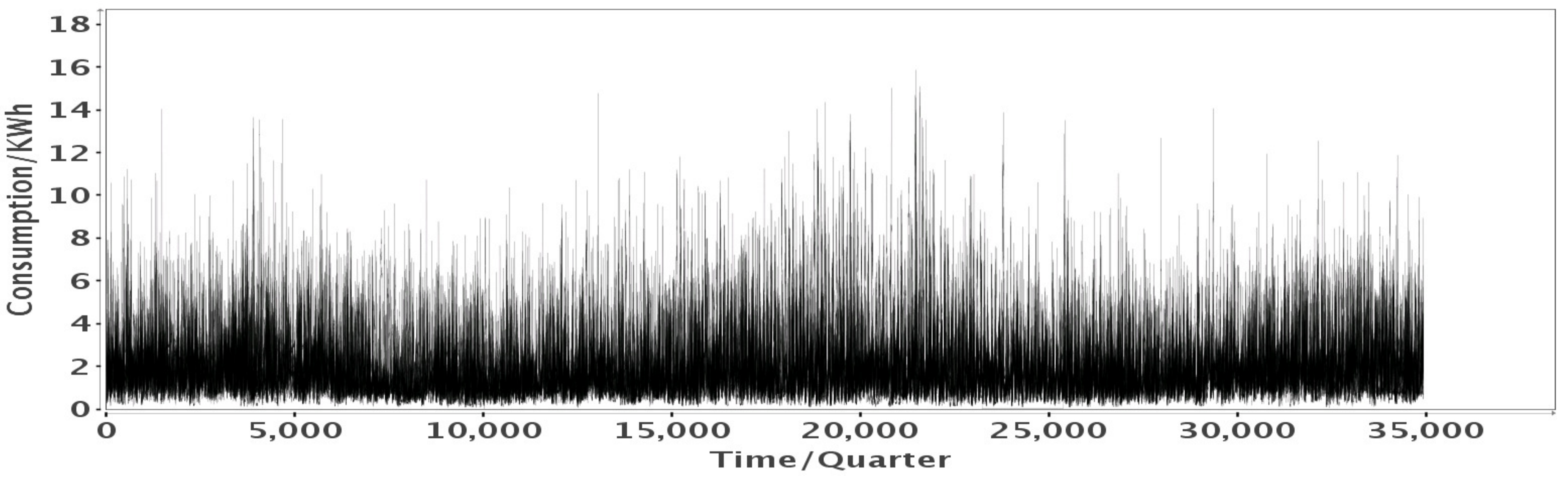}}
\subfigure[The data synthesized using the classic Markov Chain trained with the raw data.]{\includegraphics[width=2in, height=0.8in]{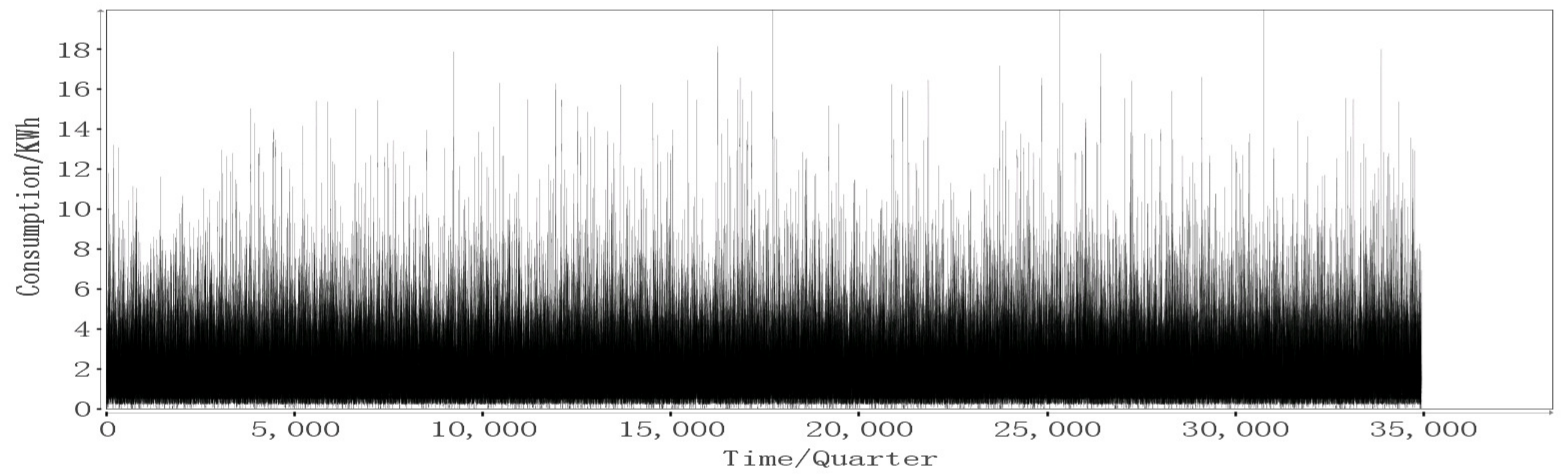}
\includegraphics[width=2in, height=0.8in]{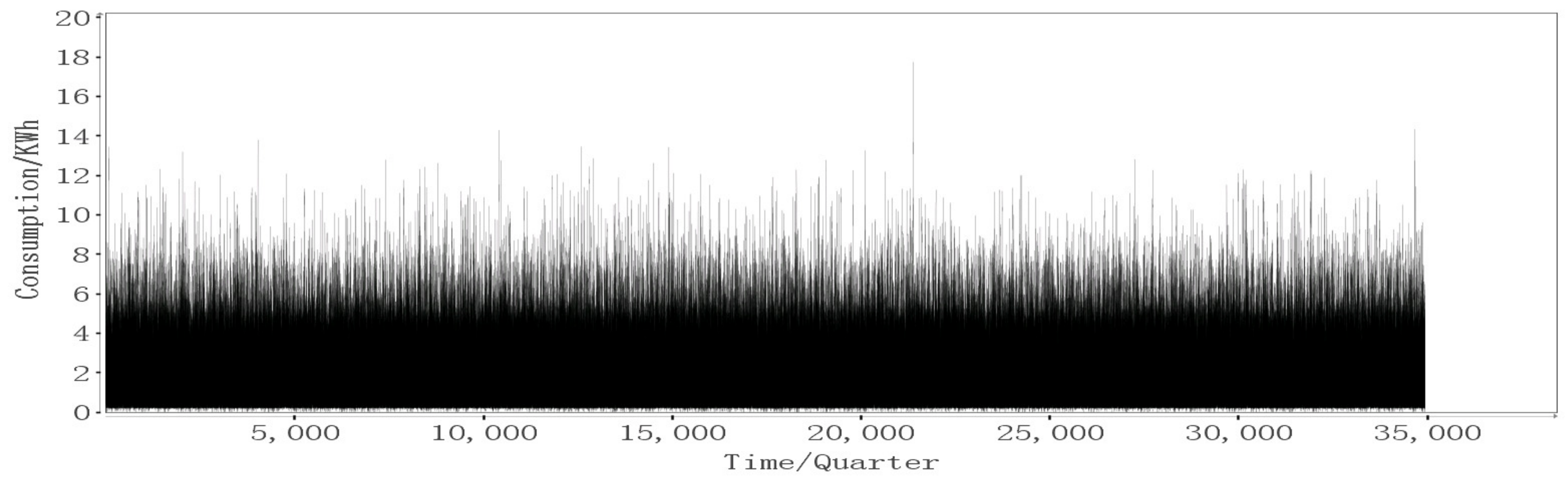}\includegraphics[width=2in, height=0.8in]{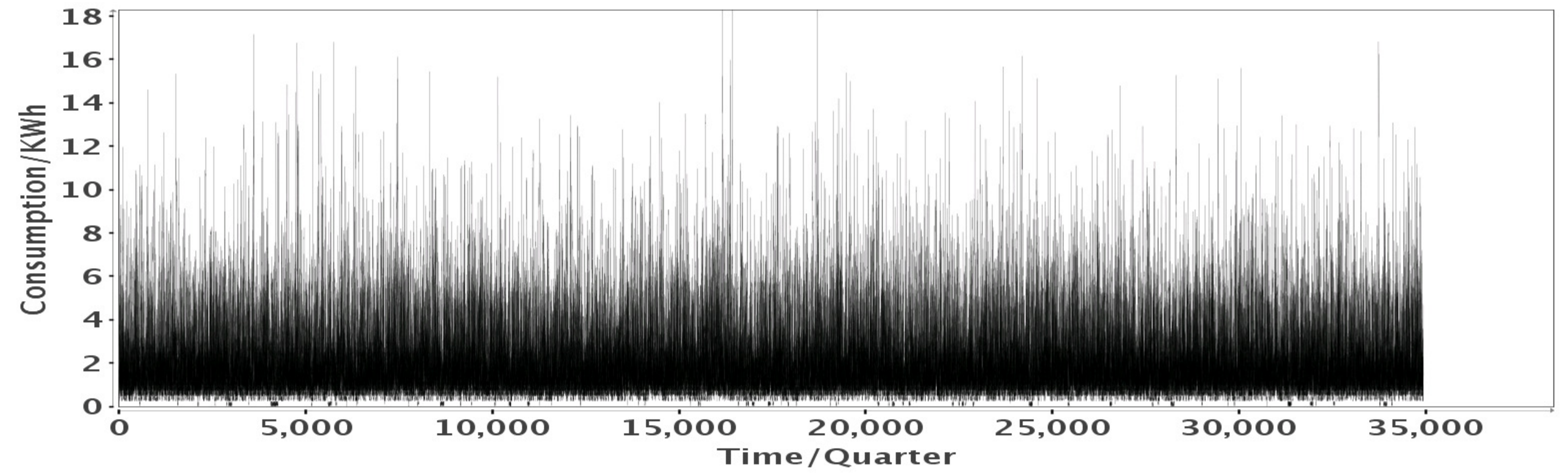}}
\caption{The raw data and synthesized yearly patterns. The left of (a) is $DS_{r\_1}$. The middle of (a) is $DS_{r\_2}$. The right of (a) is $DS_{r\_3}$. The left of (b) is $DS_{h\_1}$. The middle of (b) is $DS_{h\_2}$. The right of (b) is $DS_{h\_3}$. The left of (c) is $DS_{c\_1}$. The middle of (c) is $DS_{c\_2}$. The right of (c) is $DS_{c\_3}$.\label{Fig10}}
\end{figure*}

\begin{figure*}[t]
\centering
\subfigure[The raw data.]{\includegraphics[width=2in, height=0.8in]{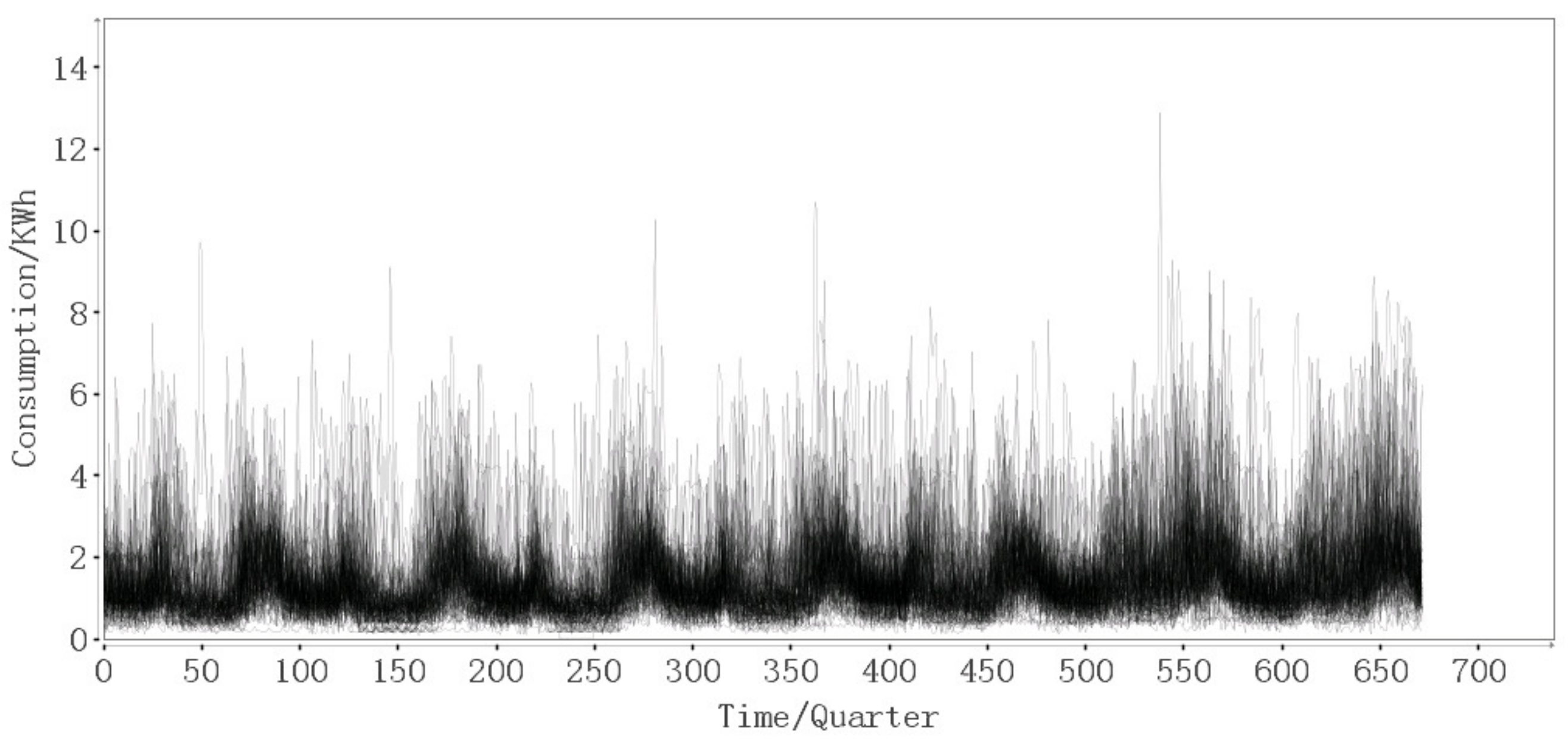}
\includegraphics[width=2in, height=0.8in]{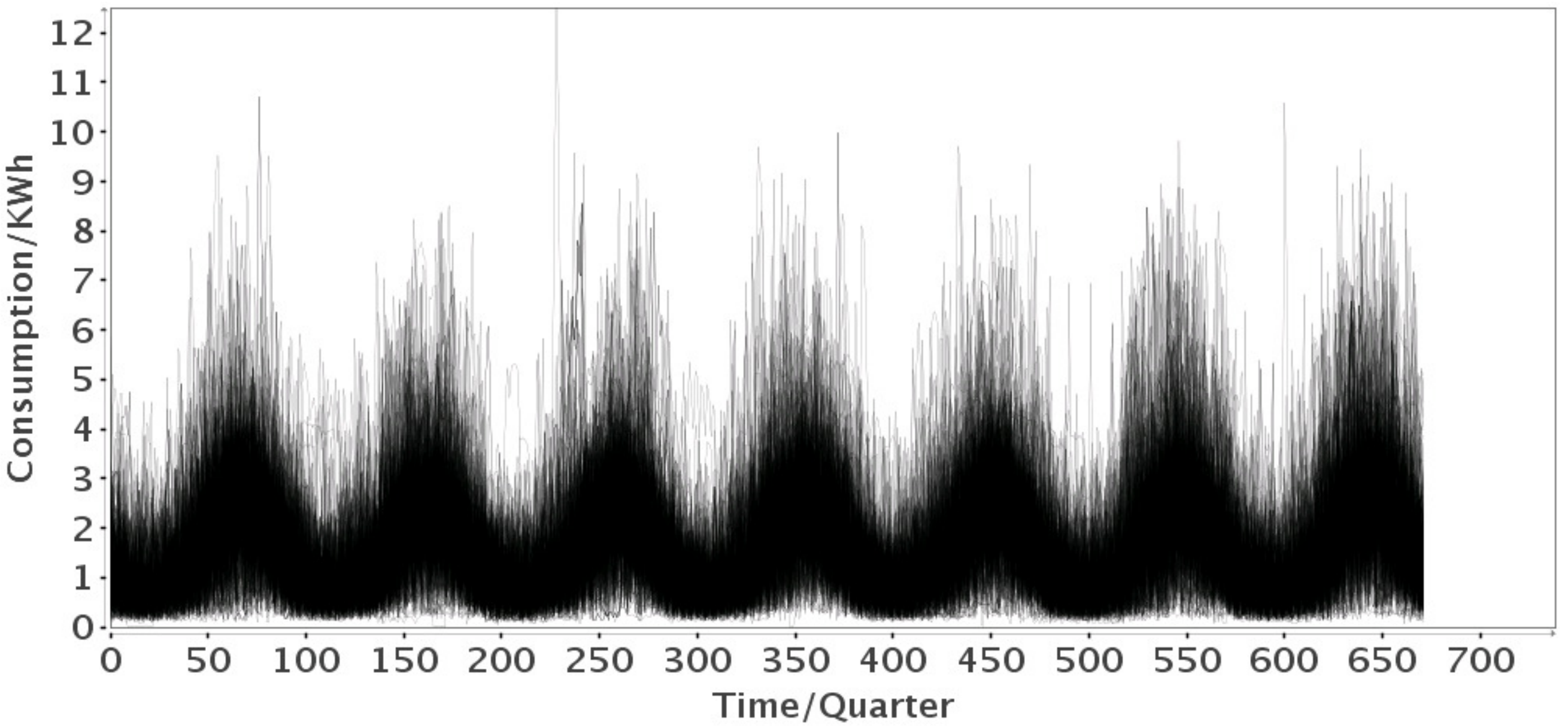}\includegraphics[width=2in, height=0.8in]{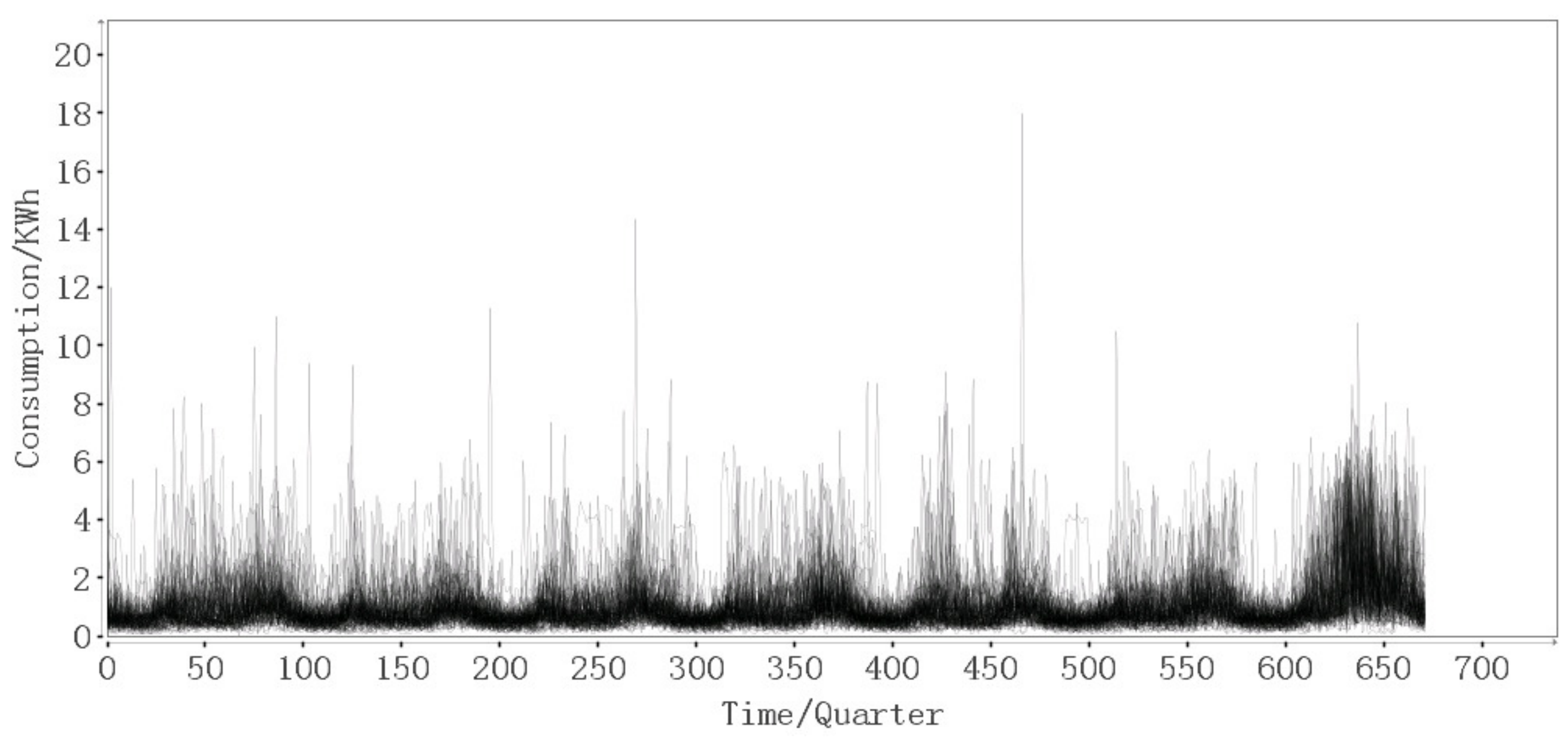}}
\subfigure[The data synthesized using the  HMMC model trained with the raw data.]{\includegraphics[width=2in, height=0.8in]{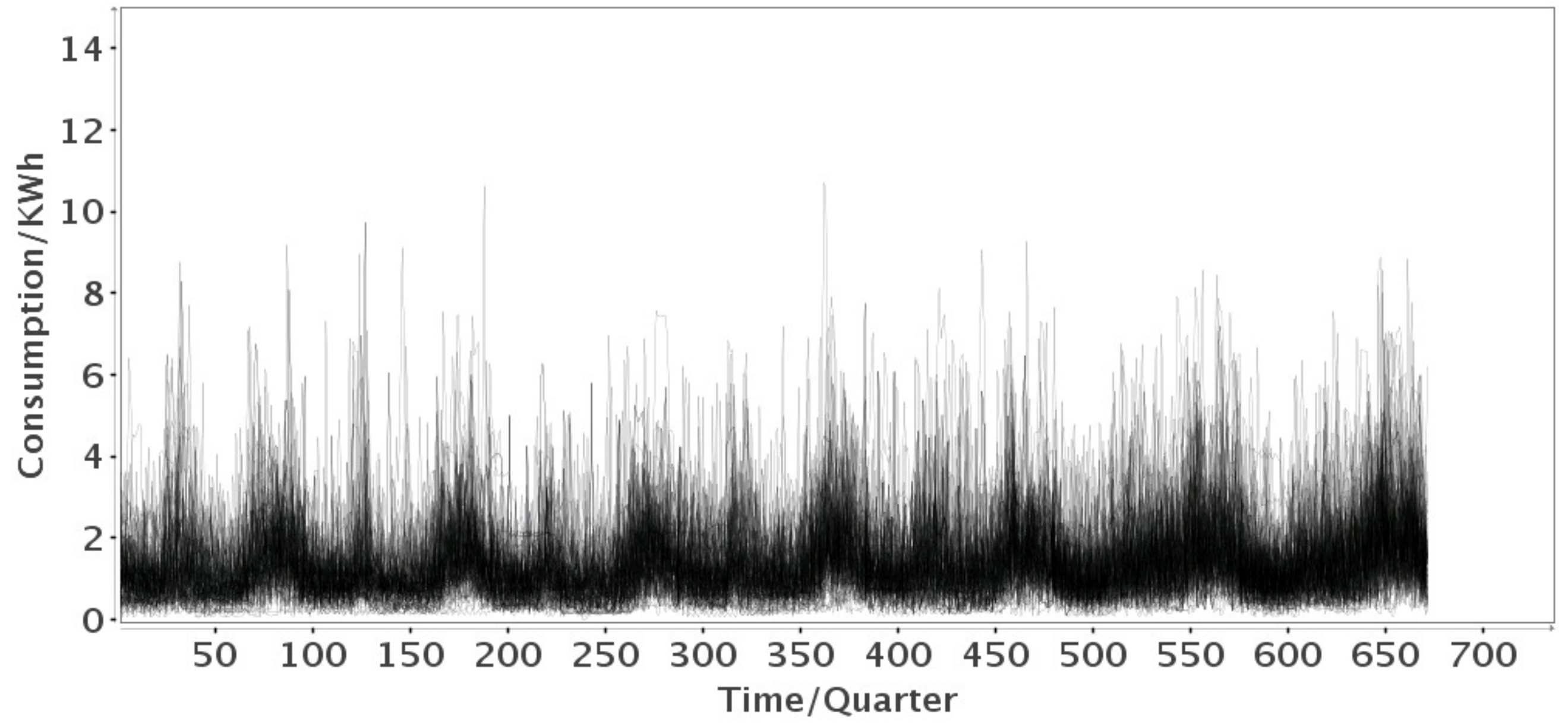}\includegraphics[width=2in, height=0.8in]{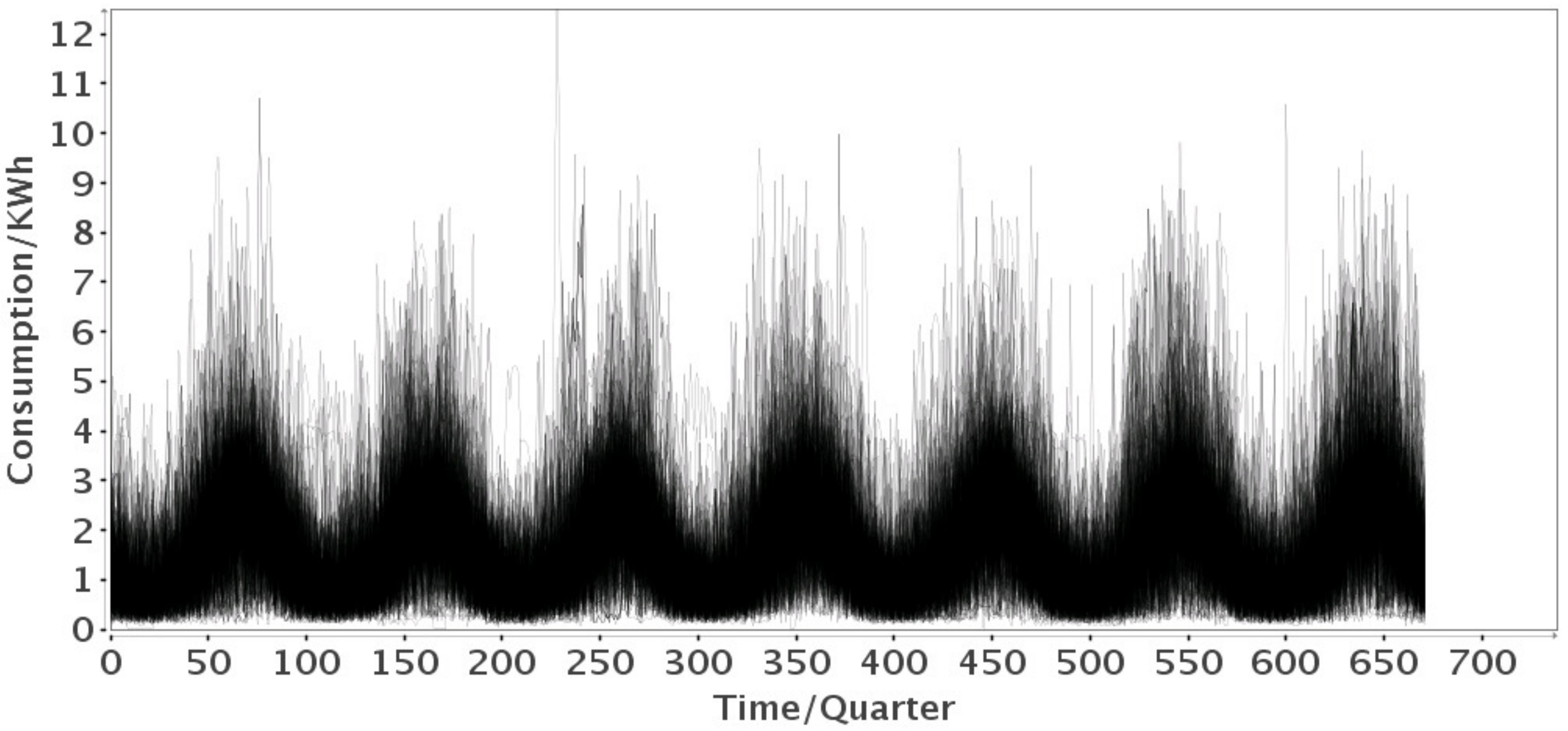}\includegraphics[width=2in, height=0.8in]{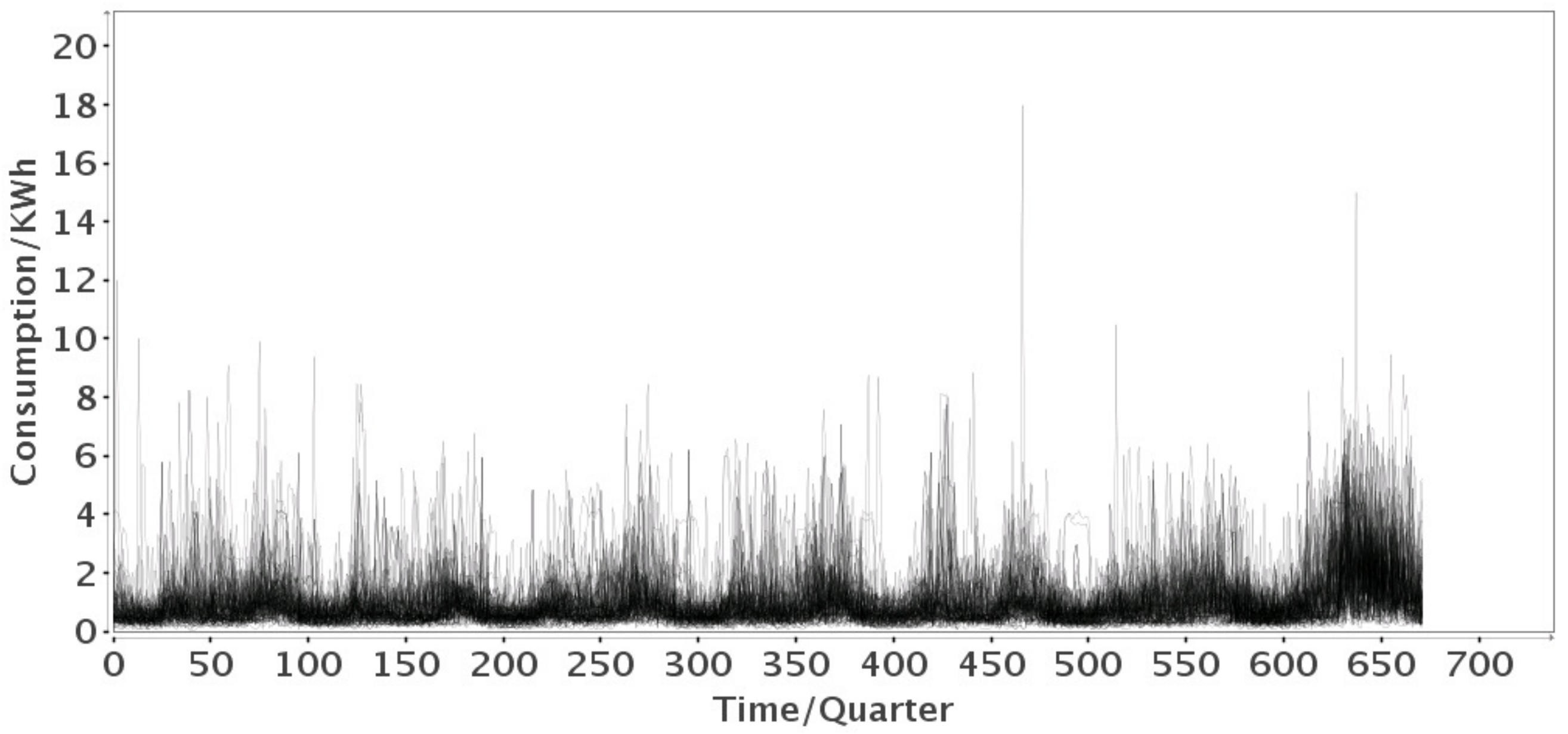}}
\caption{The raw data and synthesized weekly patterns. The left of (a) is $DS_{r\_w\_1}$. The middle of (a) is $DS_{r\_w\_2}$. The right of (a) is $DS_{r\_w\_3}$. The left of (b) is $DS_{h\_w\_1}$. The middle of (b) is $DS_{h\_w\_2}$. The right of (b) is $DS_{h\_w\_3}$.\label{Fig11}}
\end{figure*}
				
\begin{figure*}[t]
\centering
\subfigure[The raw data.]{\includegraphics[width=2in, height=0.8in]{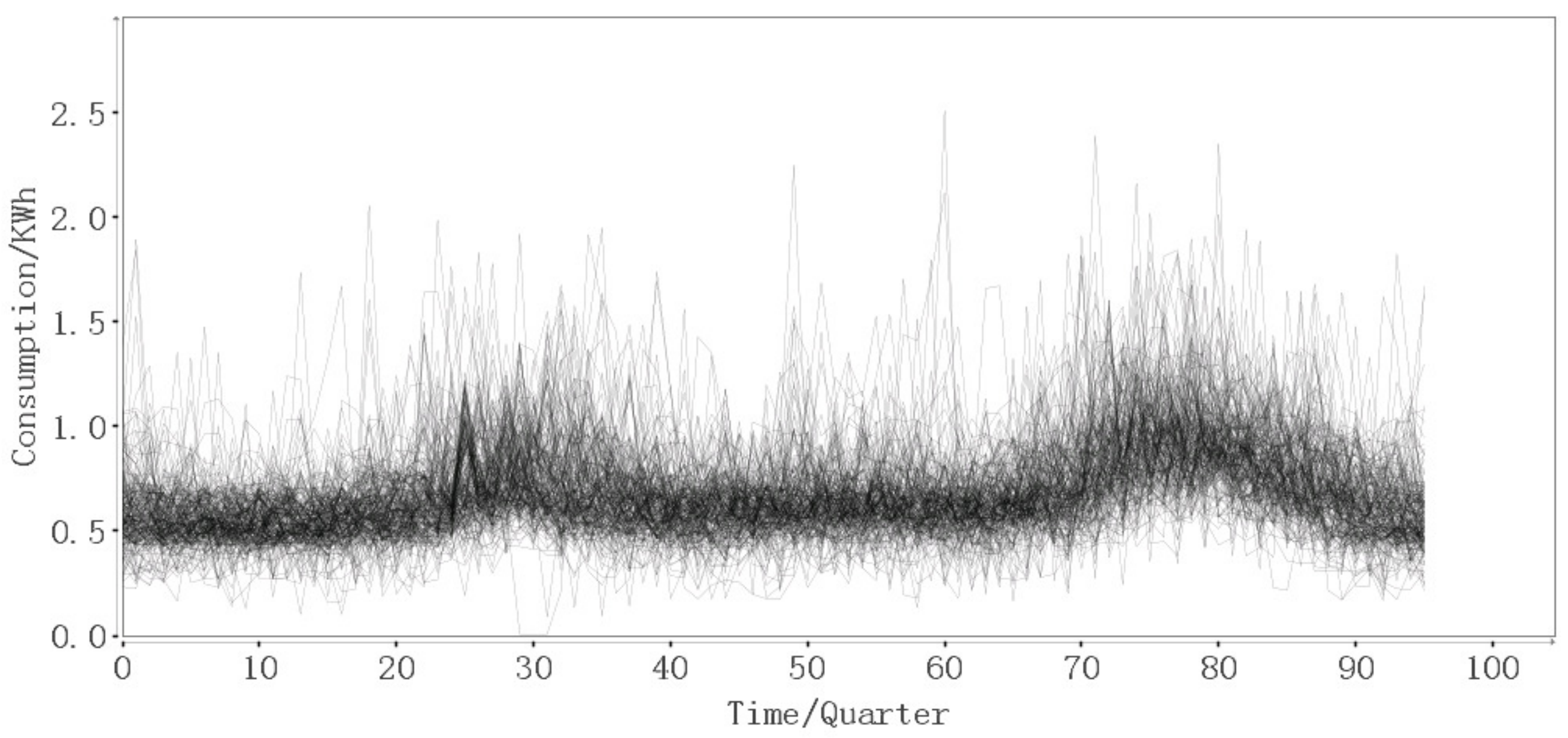}
\includegraphics[width=2in, height=0.8in]{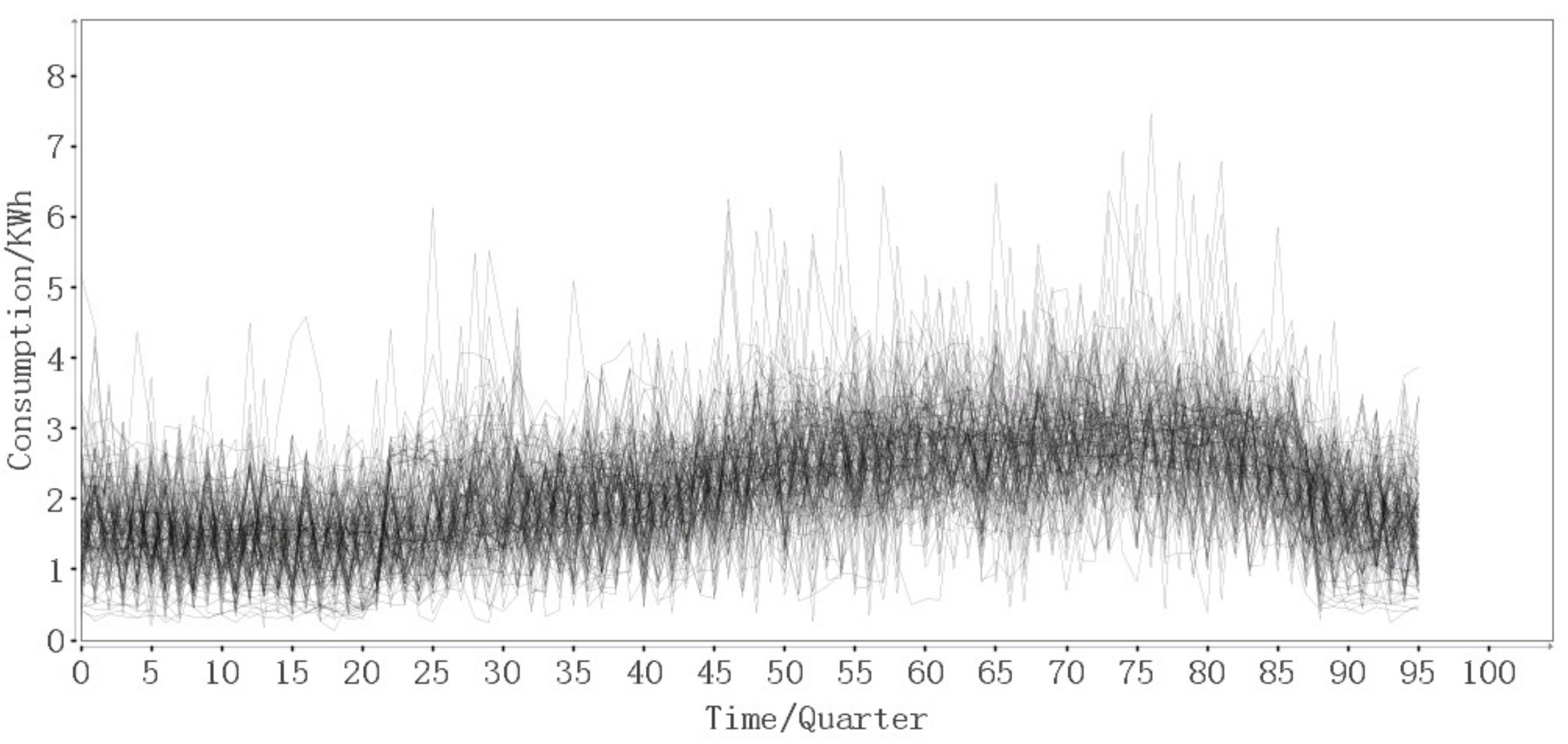}\includegraphics[width=2in, height=0.8in]{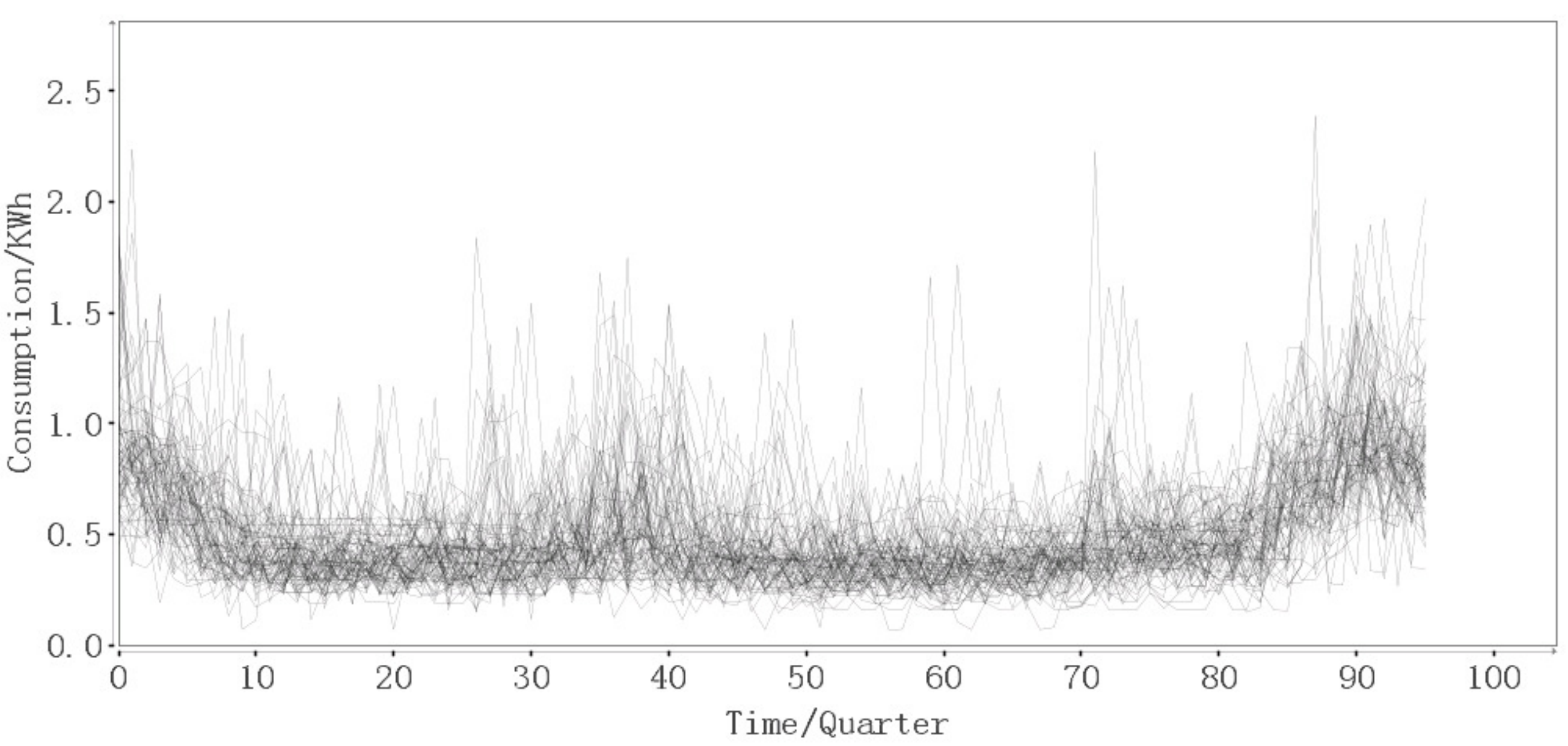}}
\subfigure[The data synthesized using the  HMMC model trained with raw data.]{\includegraphics[width=2in, height=0.8in]{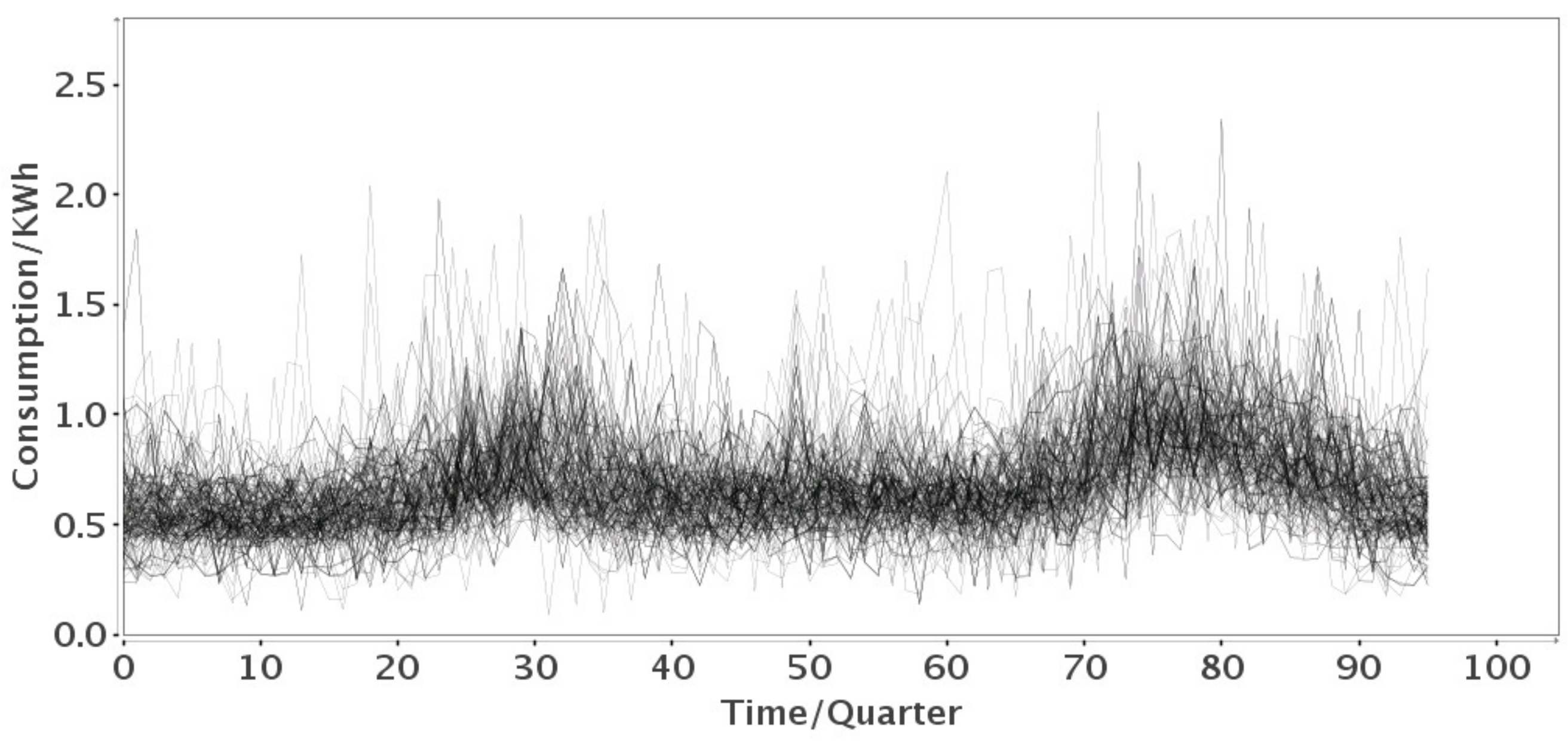}
\includegraphics[width=2in, height=0.8in]{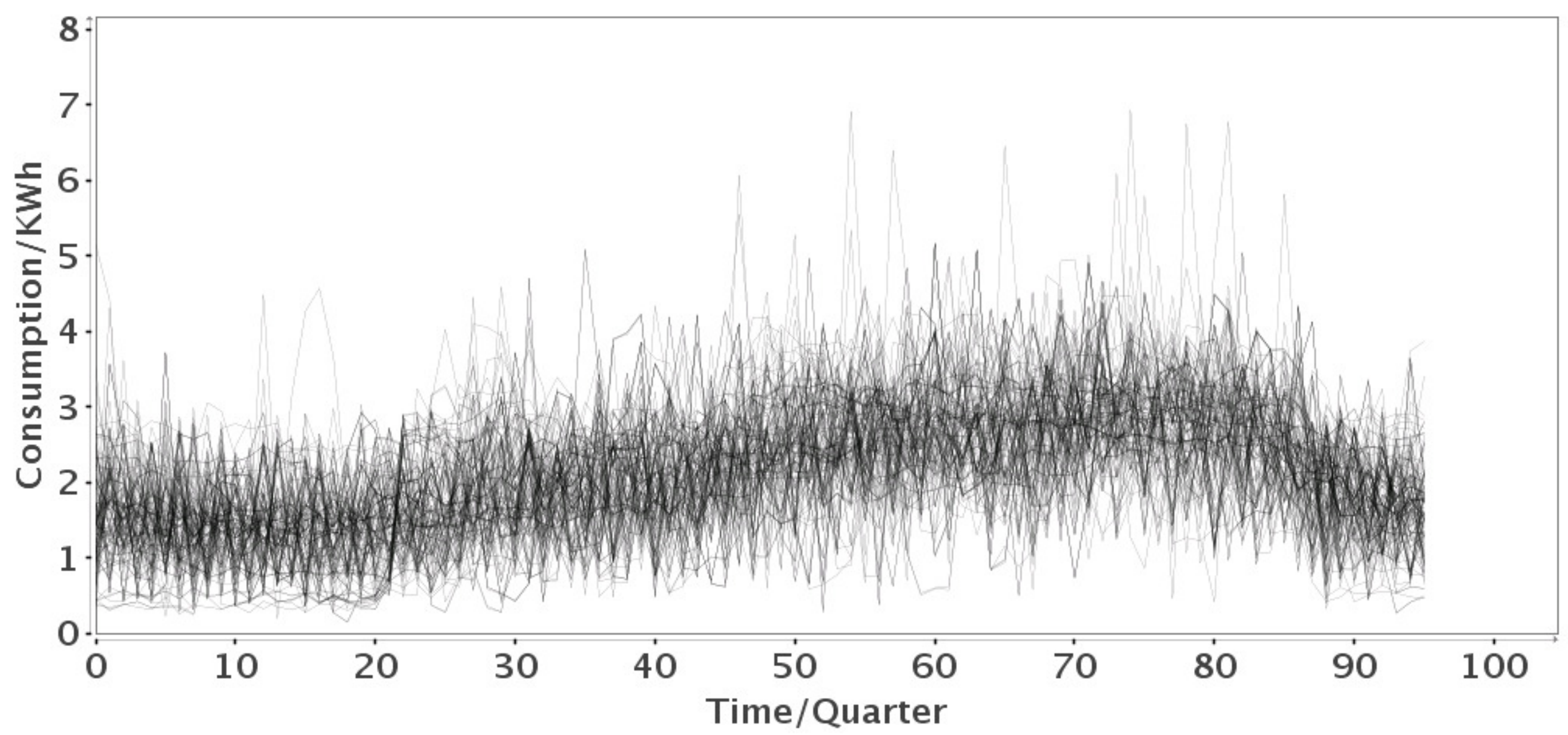}\includegraphics[width=2in, height=0.8in]{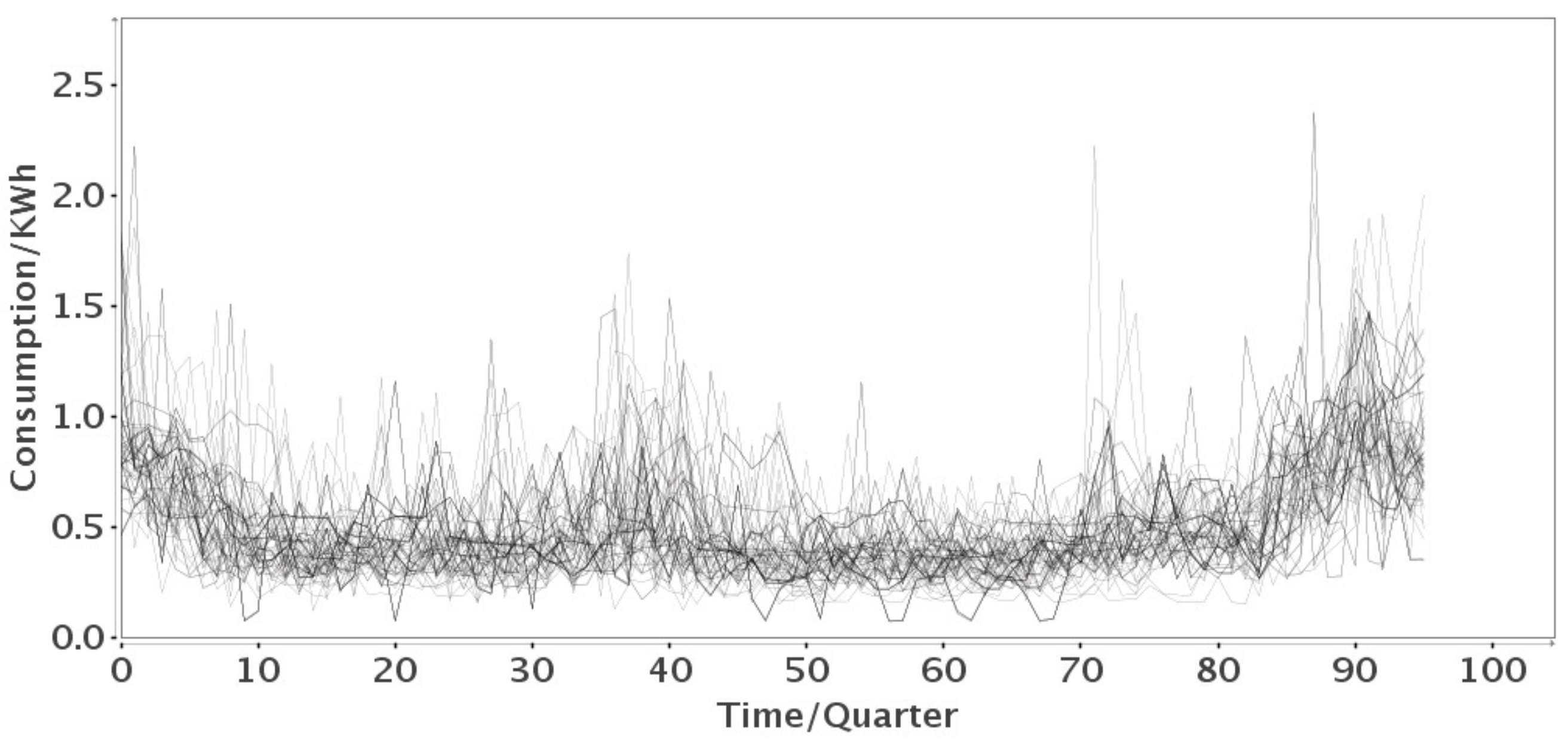}}
\caption{The raw data and synthesized daily patterns. The left of (a) is $DS_{r\_d\_1}$. The middle of (a) is $DS_{r\_d\_2}$. The right of (a) is $DS_{r\_d\_3}$. The left of (b) is $DS_{h\_d\_1}$. The middle of (b) is $DS_{h\_d\_2}$. The right of (b) is $DS_{h\_d\_3}$.\label{Fig12}}
\end{figure*}

To solve the problems mentioned above, we propose an HMMC model, which is separated into three time scales: yearly, weekly, and daily. First, as Figure~\ref{Fig7} shows, a bottom-up approach is used to model yearly electricity consumption patterns. Second,  the HMMC model in a hierarchical structure is used to model the yearly pattern. Due to the fact that a yearly pattern consists of only 52 elements---each of which represents a weekly pattern, a yearly pattern can be model by an MMC with 51 transition probability matrices on the scale of year. Likewise, every week pattern consists of only 7 elements, each of which represents a daily pattern, so it can be model by an MMC with 6 transition probability matrices on the scale of week. On the scale of day, every  daily pattern consist of only 96 elements, each of which represents the electricity consumption data, and it can be model by an MMC with 95 transition probability matrices. Modeling the yearly pattern using the HMMC model can effectively reduce the size of transition probability matrix.


Instead of directly synthesizing a yearly electricity local profile, the HMMC model firstly generate the pattern sequences on different time scales. Therefore, it ensures that each part of the synthesized electricity load profile is reasonable on the scales of  day, week, and year, respectively. As Figure \ref{Fig8} shows, a yearly electricity load profile is synthesized using the HMMC model. First, we select a yearly pattern, and synthesize the weekly pattern sequences accordingly. Second, for every weekly pattern in the sequence, we generate the daily pattern sequences accordingly. And then, for every daily pattern, we synthesize the electricity load profile. Finally, a yearly electricity load profile is generated by concatenating the daily electricity load profiles in a chronological order.

\subsection{The user model}
The user model is a stochastic one, derived from the statistics of the user information. The anonymous user information is one of the most important parts for the public dataset.  The multi-nominal logistic regression is used to calculate the likelihood of the user for a particular yearly pattern\cite{14Mcloughlin2015A}. According to Equation~\ref{eq2}, the multi-nominal logistic regression is expressed as
\begin{equation}\label{eq2}
\begin{aligned}
  logit(X)
  &=\log{\frac{p}{1-p}}\\
  &=\beta_0+\beta_1x_1+\beta_2x_2+...+\beta_nx_n
\end{aligned}
\end{equation}
Here, $X$ is an $n$ dimensions variable, which represents a user. Variables $x_1,x_2,...,x_n$ are the $n$ attributes of the user. The $\beta_0$ is a constant, and $\beta_1,\beta_2,...,\beta_n$ are the regression coefficients. $logit(X)$ is the likelihood of $X$ matching a particular yearly pattern.

\section{Results and discussion}
In the following section, the electricity consumption pattern is extracted by the clustering algorithms described in Section . For comparison, the electricity load profiles are modeled by our HMMC model and the first-order Markov Chain, respectively. 

\subsection{Extracting electricity consumption patterns}
For electricity users, the electricity consumption behavior is very variable. In general, clustering electricity load profiles would result in either massive groups or huge variances within a group. We set $ \gamma=10\%$ after making a tradeoff between the cluster size and the variances.

The adaptive K-Means algorithm is applied to the data set on different scales, respectively. Due to the variability  of the electricity consumption behavior, we get 22550 clusters in terms of daily patterns. Among them, 12155 clusters only have 1 daily electricity load profile. Likewise, we get 4013 clusters in terms of weekly patterns and 142 clusters in terms of yearly pattern. Figure~\ref{Fig9} shows typical patterns examples  of different time scales.

\subsection{Modeling the electricity consumption data}

In this paper, for every yearly pattern, we generate the electricity consumption data using the third-order HMMC model and the classic Markov model, respectively. As shown in Table~\ref{Tab1}, there are 9 data sets belonging to three yearly patterns, respectively. $DS_{r\_1}$, $DS_{r\_2}$ and $DS_{r\_3}$ are the raw data of the different yearly patterns, while $DS_{h\_1}$, $DS_{h\_2}$ and $DS_{h\_3}$ are the synthesized yearly patterns using the  HMMC model, and $DS_{c\_1}$, $DS_{c\_1}$ and $DS_{c\_1}$ are the synthesized yearly patterns using the classic Markov Chain model.

As Table~\ref{Tab1} shows, the classic Markov model can preserve partial  features of the original electricity load profiles. For example, $\mu$, $\sigma$, $p_{max}$,  $p_{min}$ and $\mu_{total}$ are very similar to the  raw data. In terms of the metrics---$\sigma_{pro}$, $\sigma_{total}$, $\gamma_{\sigma,\mu}$, $c_{max}$ and $c_{min}$, the HMMC model performs much better than the classic Markov Chain model in the experiments. Compared with the raw data,  $\sigma_{pro}$, $\sigma_{total}$, $\gamma_{\sigma,\mu}$, $c_{max}$ and $c_{min}$ are  more similar using the HMMC model than using the classic Markov Chain model.  And Figure~\ref{Fig10} shows, comparing the data synthesized using the classic Markov Chain model with the raw data, there are significant differences between the raw data and the synthesized one. We notice that there are  unreasonable parts in the electricity load profiles synthesized with the classic Markov Chain, while the raw data is very similar to the data synthesized using the HMMC model. The reason is that the differences between the electricity load profiles within the pattern is accumulated when electricity load profiles are synthesized  using  the classic Markov Chain.

\begin{table*}[t]
\scriptsize
\centering
\begin{threeparttable}
\caption{The metrics of the raw data and synthesized data of the yearly pattern.\label{Tab1}}
\begin{tabular}{ccccccccccc}
\hline
 $DS$&$\mu$ &$\sigma$ &$p_{max}$ &$p_{min}$ & $\sigma_{pro}$ &$\mu_{total}$ &$\sigma_{total}$ & $\gamma_{\sigma,\mu}$ & $c_{max}$ &$c_{min}$\\ \hline
$DS_{r\_1}$ &1.7081&283.4584&18.1914&0&17.3386&59687.8133&5622.8101&0.0942&72038.6322&50071.5674\\
$DS_{h\_1}$ &1.7066&289.7538&24.2124&0&18.1202&59636.9444&5523.9014&0.0926&74252.8144&49215.2737\\
$DS_{c\_1}$ &1.7084&283.7152&18.1909&0&20.7554&59697.8092&934.0054&0.0156&62602.77346&56469.2666\\

$DS_{r\_2}$ &1.8641&283.5548&25.2317&0&21.2238&65140.3193&5628.9001&0.0864&74914.6572&56859.6346\\
$DS_{h\_2}$ &1.8432&280.1117&27.0431&0&20.0419&64407.6467&5561.7444&0.0864&75686.8546&54852.7722\\
$DS_{c\_2}$ &1.8629&284.0796&25.2308&0&19.8110&65098.3931&730.0663&0.0112&67690.3262&62424.4531\\

$DS_{r\_3}$ &1.8343&273.9289&23.1533&0&23.1950&64098.1363&5755.0841&0.0898&73673.1285&58454.8488\\
$DS_{h\_3}$ &1.8203&273.7401&23.1533&0&21.7358&63608.9406&5639.3368&0.0887&74023.4893&57622.6536\\
$DS_{c\_3}$ &1.8339&274.4035&23.1526&0&20.1366&64083.5409&827.8379&0.0129&66674.0676&61872.8854\\
\hline
\end{tabular}
\begin{tablenotes}
    \item[1] Here, the metrics include: $\mu$, $\sigma$,  $p_{max}$, $p_{min}$, $\sigma_{pro}$, $\mu_{total}$, $\sigma_{total}$, $\gamma_{\sigma,\mu}$, $c_{max}$ and $c_{min}$. $\mu=\frac{\mu_{total}}{d}$ $d$ is the dimension of the electricity load profile, $\mu_{total}$ is the mean of the total consumption. $\sigma$ is an metric that measures  the variance of electricity consumption.  $p_{max}$ is the highest consumption of the electricity load profile and $p_{min}$ is the smallest consumption of the electricity load profile.  $\sigma_{pro}$  is a metric that measures  the variance of the electricity load profiles within a pattern. $\sigma_{pro}=\sqrt{\|X\|}$, $X=\{x_1,x_2,...,x_d\}$, $x_i$ is the variance of $ith$ dimension within a  pattern. $\sigma_{total}$ is the standard deviation of the total consumption. $\gamma_{\sigma,\mu}=\frac{\sigma_{total}}{\mu_{total}}$ . $c_{max}$ is the biggest total consumption. $c_{min}$ is the smallest total consumption.
\end{tablenotes}
\end{threeparttable}
\end{table*}

\begin{table*}[t]
\scriptsize
\centering
\begin{threeparttable}
\caption{The metrics of raw data and synthesized weekly and daily patterns.\label{Tab2}}
\begin{tabular}{cccccccccccc}
\hline
 $DS$&$\mu$ &$\sigma$ &$p_{max}$ &$p_{min}$ & $\sigma_{pro}$&$\mu_{total}$ &$\sigma_{total}$ &$\gamma_{\sigma,\mu}$ & $c_{max}$ &$c_{min}$\\ \hline
$DS_{r\_w\_1}$ &1.3376&22.7966&12.8959&0&4.3966&898.8339&68.0509&0.0757&1130.3786&754.0151\\
$DS_{h\_w\_1}$ &1.3289&24.3616&12.8956&0&4.6772&892.9220&75.5433&0.0846&1232.0509&715.8526\\
$DS_{r\_w\_2}$ &1.6015&30.1395&10.7404&0&4.9419&1076.2008&83.2836&0.0774&1261.0420&875.3925\\
$DS_{h\_w\_2}$ &1.6088&30.7867&17.8931&0&5.2270&1081.0966&93.0269&0.0860&1343.5733&818.8636\\
$DS_{r\_w\_3}$ &0.8965&19.2843&17.9762&0&4.0192&602.4527&52.6302&0.0874&717.0628&485.4517\\
$DS_{h\_w\_3}$ &0.9017&20.1068&17.9762&0&4.0539&605.9462&60.4593&0.0997&804.7128&459.7600\\

$DS_{r\_d\_1}$ &0.6742&2.0997&2.5089&0.0070&0.6003&64.7274&3.4618&0.0535&71.4993&57.5900\\
$DS_{h\_d\_1}$ &0.6744&2.0978&2.5088&0.0071&0.5986&64.7439&3.4515&0.0533&72.6561&56.2700\\
$DS_{r\_d\_2}$ &0.5034&2.4070&2.3880&0.0690&0.6355&48.3255&3.3699&0.0697&56.2327&41.1594\\
$DS_{h\_d\_2}$ &0.5041&2.4114&2.3879&0.0690&0.6363&48.3937&3.4515&0.0713&56.3205&41.1317\\
$DS_{r\_d\_3}$ &2.1439&8.4492&7.4637&0.1314&2.1745&205.8188&12.1681&0.0591&239.0681&181.9523\\
$DS_{h\_d\_3}$ &2.1449&8.4412&7.4634&0.1315&2.1709&205.9077&12.0770&0.0587&239.4820&181.6977\\
\hline
\end{tabular}
\end{threeparttable}
\end{table*}

Besides, the HMMC model preserves the real features of electricity load profiles of the scales of week and day. Table~\ref{Tab2} shows 12 data sets belonged to 6 different patterns, respectively. $DS_{r\_w\_1}$, $DS_{r\_w\_2}$ and $DS_{r\_w\_3}$ are the raw data of different weekly patterns. Accordingly, $DS_{h\_w\_1}$, $DS_{h\_w\_2}$, and $DS_{h\_w\_3}$ are the synthesized weekly patterns using the HMMC model. Similarly, $DS_{r\_d\_1}$, $DS_{r\_d\_2}$ and $DS_{h\_d\_3}$ are the raw data of different daily patterns, and $DS_{h\_d\_1}$, $DS_{h\_d\_2}$, and $DS_{h\_d\_3}$ are synthesized daily patterns with the HMMC model.

As shown in Table~\ref{Tab2}, the synthesized  data is similar to the raw data of the scales
of week and day. Figure~\ref{Fig11} shows, the synthesized weekly patterns using the HMMC model are similar to the raw electricity load  profiles. Likewise, Figure~\ref{Fig12} also  shows, the synthesized daily patterns are similar to the raw  electricity load profiles.
On the overall, the HMMC model preserves the real electricity consumption behavior of different time scales.

\subsection{User information}

There are several sensitive information in the user information. We filter most of the sensitive information, and only keep non-sensitive information. 
With the probabilistic model derived from the statistics of the user information, we synthesize  the base user information. And then we assign a yearly consumption pattern to a user according to the likelihood of the user matching a particular yearly pattern.

\section{Public models}
The real data set is the key of the data synthesis. We provide two models on the basis of two real data sets, respectively. The first one is a public data set---Pecan Street Inc Dataport2017~\cite{22PecanStreetInc}, and it contains 711 users. Each user data contains two types of data: user information, such as the type of building, the construction year of house, and total square footage; and electricity consumption data, collected every 15 minutes in 2015. The other one is a confidential data set, which we have obtained permission to use, and it contains data from 80000 users from the non-resident sectors. Each user data consists of two types of data: user information, including the installation year of electricity meter, address code, and industry code; and electricity consumption data, which is collected every 15 minutes in 2015.

\section{Conclusion}
The shortage of the electricity load profiles is a huge obstacle to the research on electricity consumption behaviors. Data synthesis is one of the best approach to tackling this obstacle. We propose a hierarchical multi-matrices Markov Chain model (HMMC) to synthesizing scalable electricity load profile that preserves the real consumption behaviors on three time scales: per day, per week, and per year. To promote the research of the electricity consumption behaviors, we use the HMMC model to characterize two distinctive raw electricity load profiles. One is collected from the resident sector, and the other is collected from the non-resident
sectors, including different industries such as education, finance, and manufacturing. We publish two trained models online, and researchers can directly use these trained models to synthesize scalable electricity load profiles for further researches.

\section*{Acknowledgements}
We are very grateful to anonymous reviewers. This work is supported by the Major Program of National Natural Science Foundation of China (Grant No. 61432006), National Key Research and Development Program of China (2016YFB1000600, 2016YFB1000601).



\end{document}